\documentclass{aa}  
\usepackage{natbib}
\bibpunct{(}{)}{;}{a}{}{,}
\usepackage{graphicx}
\usepackage{enumitem}
\usepackage{epstopdf}
\usepackage[percent]{overpic}
\usepackage{color}

\begin{document}
\authorrunning{Chatzistergos et al.}
\titlerunning{Segmentation of Ca~II~K SHG}
\title{Analysis of full disc Ca~II~K spectroheliograms \\II. Towards an accurate assessment of long-term variations in plage areas}
\author{Theodosios Chatzistergos\inst{1,2}, Ilaria Ermolli\inst{1},  Natalie A. Krivova\inst{2}, Sami K. Solanki\inst{2,3}}
\offprints{Theodosios Chatzistergos  \email{chatzistergos@mps.mpg.de, theodosios.chatzistergos@inaf.it}}
\institute{INAF Osservatorio Astronomico di Roma, Via Frascati 33, 00078 Monte Porzio Catone, Italy \and Max Planck Institute for Solar System Research, Justus-von-Liebig-weg 3,	37077 G\"{o}ttingen, Germany \and School of Space Research, Kyung Hee University, Yongin, Gyeonggi 446-701, Republic of Korea}
\date{}
	
\abstract
{Reconstructions of past irradiance variations require suitable data on solar activity. The longest direct proxy is the sunspot number, and it has been most widely employed for this purpose. These data, however, only  provide  information on the surface magnetic field emerging in sunspots, while a suitable proxy of the evolution of the bright magnetic features, specifically faculae/plage and network, is missing. This information can potentially be extracted from the historical full-disc observations in the Ca~II~K line.}
{We use several historical archives of full-disc Ca~II~K observations to derive plage areas over more than a century. Employment of different datasets allows identification of systematic effects in the images, such as changes in instruments and procedures, as well as an assessment of the uncertainties in the results.} 
{We have analysed over 100,000 historical images from 8 digitised photographic archives of the Arcetri, Kodaikanal, McMath-Hulbert, Meudon, Mitaka, Mt Wilson, Schauinsland, and Wendelstein observatories, as well as one archive of modern observations from the Rome/PSPT. The analysed data cover the period 1893--2018. We first performed careful photometric calibration and compensation for the centre-to-limb variation, and then segmented the images to identify plage regions. This has been consistently applied to both historical and modern observations. } 
{The plage series derived from different archives are generally in good agreement with each other. However, there are also clear deviations that most likely hint at intrinsic differences in the data and their digitisation. We showed that accurate image processing significantly reduces errors in the plage area estimates. Accurate photometric calibration also allows precise plage identification on images from different archives without the need to arbitrarily adjust the segmentation parameters. Finally, by comparing the plage area series from the various records, we found the conversion laws between them. This allowed us to produce a preliminary composite of the plage areas obtained from all the datasets studied here. This is a first step towards an accurate assessment of the long-term variation of plage regions.}

	
\maketitle
	
\section{Introduction}
\label{sec:intro}
\sloppy

The Sun is the main external energy supplier to the Earth.
The radiative energy from the Sun received per unit time and area at 1 AU is termed solar irradiance. Thus knowledge of solar irradiance is of crucial interest for climate studies \citep[][]{haigh_sun_2007,gray_solar_2010,ermolli_recent_2013,solanki_solar_2013-1}. 
Direct measurements of solar irradiance are available since 1978. They show that the total solar irradiance (i.e., integrated over the entire solar spectrum) varies by about 0.1\% in phase with the solar activity cycle, with sometimes even stronger fluctuations on the time scales of days. The records of direct measurements are however too short to investigate the solar impact on the complex climate system of Earth properly. This calls for sufficiently long reconstructions of past irradiance variations.

Solar irradiance changes on time scales of days and longer are generally attributed to varying contributions of dark sunspots and  bright facular and network regions \citep{shapiro_nature_2017}. 
Models assuming that irradiance variability is driven by the solar surface magnetic flux reproduce over 90\% of the measured total solar irradiance variations \citep[see e.g.][and references therein]{yeo_reconstruction_2014,yeo_solar_2017}.
The success of these models lent grounds to believe that reconstructions of past irradiance variability can give a realistic picture of the Sun's behaviour, provided there are suitable data of solar magnetic activity.
Direct measurements of the solar surface magnetic field exist only for roughly the same period as covered by irradiance measurements. To go further back in time, sunspot observations and isotope data are the typically employed proxies of solar magnetic activity \citep[e.g.][]{lean_reconstruction_1995,solanki_reconstruction_1999,wang_modeling_2005,steinhilber_total_2009,krivova_reconstruction_2007,krivova_reconstruction_2010,delaygue_antarctic_2011,shapiro_new_2011,vieira_evolution_2011,dasi-espuig_modelling_2014,dasi-espuig_reconstruction_2016,cristaldi_1d_2017,wu_solar_2018}. However, sunspot observations do not carry information on faculae or the network, whereas isotope data trace primarily the open magnetic flux, which is only indirectly related to sunspot and facular fields. 
These limitations can be overcome by combining observations taken in white-light and in the Ca~II~K  spectral band to monitor sunspots and facular regions, respectively. Such observations cover a period exceeding a century. 
Owing to the tight relationship between the Ca~II~K brightness and the non-sunspot  magnetic field strength averaged over a pixel \citep[][]{babcock_suns_1955,skumanich_statistical_1975,schrijver_relations_1989,ortiz_how_2005,loukitcheva_relationship_2009,pevtsov_reconstructing_2016,chatzistergos_analysis_2017,kahil_brightness_2017}, the Ca~II~K observations can provide detailed information on the surface coverage by bright faculae and network regions (hereafter referred to as plage and network as Ca~II~K observations trace the chromosphere, where facular regions are called plage). This information should allow a significant improvement of long-term irradiance models, at least for the period of time that Ca~II~K data are available and possibly, indirectly, also for even longer periods of time. 

\begin{figure*}[t!]   \centering 
	\includegraphics[width=1\linewidth]{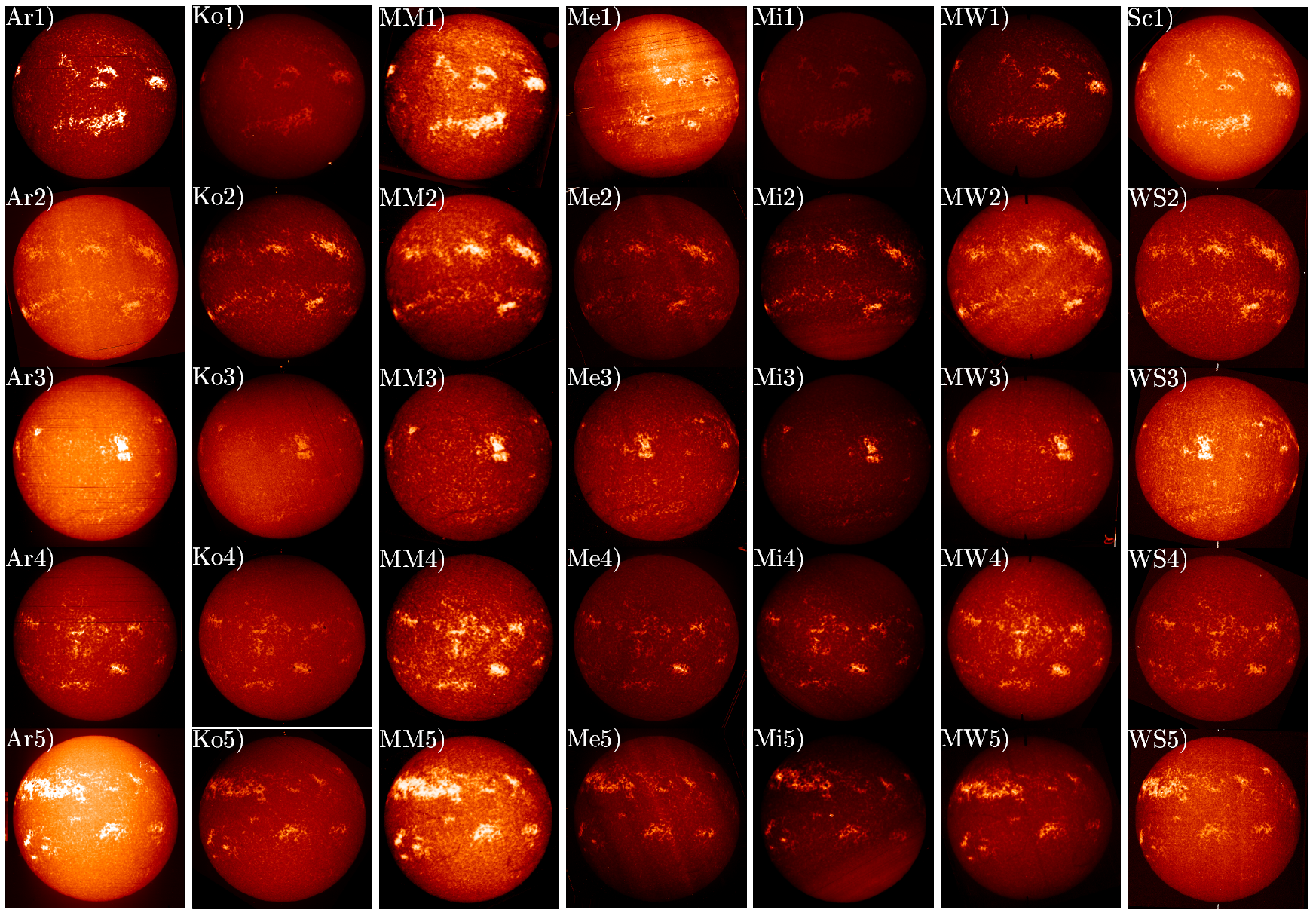}
	\caption{Raw Ca II K density images from the Arcetri (Ar), Kodaikanal (Ko), McMath-Hulbert (MM), Meudon (Me), Mitaka (Mi), Mt Wilson (MW), Schauinsland (Sc), and Wendelstein (WS) observatories (Sc and WS are shown in a single column due to their low number of images). Each row shows observations from the different archives taken on roughly the same day, except for Me1. The dates of the observations are (from top to bottom): 09/12/1959 (17/11/1893 for Me), 01/03/1968 (29/02/1968 for Me), 01/08/1968 (30/07/1968 for Me and WS), 05/09/1968 (06/09/1968 for WS), and 18/03/1969 (17/03/1969 for Me and WS). All images are normalised to the range of values within the disc and are compensated for ephemeris.}
	\label{fig:processedimagessamedayoriginal} \end{figure*}

In recent years, digitization of various archives \citep[for a list of the available archives see][]{chatzistergos_analysis_2017} allowed starting extensive exploitation of historical Ca~II~K spectroheliograms \citep[SHG, e.g.][]{ribes_search_1985,kariyappa_contribution_1996,foukal_behavior_1996,foukal_extension_1998,caccin_variations_1998,worden_evolution_1998,zharkova_full-disk_2003,lefebvre_solar_2005,ermolli_comparison_2009,ermolli_digitized_2009,tlatov_new_2009,bertello_mount_2010,dorotovic_north-south_2010,sheeley_carrington_2011,priyal_long_2014,priyal_long-term_2017,chatterjee_butterfly_2016,chatterjee_variation_2017}. 
Overall, there are numerous published results from historical Ca~II~K images, which agree in some respects, but also show significant differences \citep[see][]{ermolli_potential_2018}.
A critical aspect of most of these studies is, however, that they used photometrically uncalibrated images and did not analyse the possible systematic changes of the quality of the images recorded by different observers over decades \citep[see the discussion in][]{ermolli_comparison_2009}. For example, simple visual examination of the available data shows that co-temporal observations from different series can diverge significantly due to the differences in spectral and spatial resolution of observations, stray-light, noise, calibration, and other instrumental and archival characteristics. Besides, diverse processing methods were applied to the data from individual archives, thus hampering comparison of results from the various series. An accurate and homogeneous compilation of multiple datasets  into a single series, critical for irradiance reconstructions and other studies, is thus a challenging task.

In \cite{chatzistergos_analysis_2018} we introduced a novel approach to process the historical SHG observations and showed that it allows to perform the photometric calibration and account 
for image artefacts with higher accuracy than obtained with other methods presented in the literature. We also showed that the proposed method returns consistent results for images from different SHG archives \citep[for examples see also][]{chatzistergos_ca_2018}.

In this paper we analyse carefully reduce and calibrate images from 8 historical and one modern archives of Ca~II~K observations to evaluate changes in plage area coverage from 1893 to 2018. Analysis of  multiple archives helps  us to assess systematic changes and sources of artefacts in the results derived from individual datasets.

The paper is organized as follows. In Sections \ref{sec:data} and \ref{sec:preprocessing} we describe the data analysed in our study and present a brief overview of the methods applied, respectively.  Our results for the plage areas are presented and compared with other published results in Section \ref{sec:results}. In Section \ref{discussion} we discuss the effects of photometric calibration and other processing steps on the results, while in Section \ref{sec:composite} we produce preliminary composite series from the analysed data. In Section \ref{sec:conclusions} we summarize the results of our study and draw our conclusions.

\section{Data}
\label{sec:data}
We consider full-disc Ca~II~K observations derived from the digitization of 8 historical SHG photographic archives and from  the modern Rome/PSPT (Precision Solar Photometric Telescope). The latter data allow us to extend the time-series of plage areas derived from the historical observations to the present day as well as to test the accuracy of our processing.  The plage areas derived in this work are compared to other published records and to the sunspot area record compiled by \citet{balmaceda_homogeneous_2009}\footnote{Available at \url{http://www2.mps.mpg.de/projects/sun-climate/data.html}} 
\subsection{Full-disc Ca II K archives}
\label{sec:historicaldata}

The historical data analysed here are the digitized Ca~II~K full-disk SHG from the Arcetri  
(Ar), Kodaikanal (Ko), McMath-Hulbert (MM), Meudon (Me), Mitaka (Mi), Mount Wilson (MW), Schauinsland (Sc), and Wendelstein (WS)  photographic archives. 
In brief, the  
observations were taken between 1893 and 2002 in the Ca~II~K  line at $\lambda$ = 3933.67 \AA,  
with bandwidths ranging from 0.1 to 0.5 $\text{\AA}$. The original photographic observations, which were  
mainly  stored  on photographic plates (MM observations are on sheet film, while Mi observations after 02/03/1960 are on 24x35 mm sheet film), were digitized by using different devices and methods. Information about the devices and settings that were used to convert the Ar, Ko, and MW data to digital images can be found in e.g.
\citet[][and references therein]{ermolli_comparison_2009}, about Me in \citet{mein_spectroheliograms_1990}, about Mi in \citet{hanaoka_long-term_2013}, and Sc and WS in \citet{wohl_old_2005}.
Table \ref{tab:observatories} summarizes key information about the different datasets used in this study.

Some series derive from combinations of observations taken at different sites, while some underwent multiple digitisations. For example, the Mi data were obtained at two different sites, Azabu (1917-1924) and Mitaka (1925-1974), and have been digitised three times, once with 8-bit depth and twice with 16-bit depth. Furthermore, the second digitisation includes only the data prior to 02/03/1960, while the third digitisation includes the data after 02/03/1960 as well as the first 10 observations of each year prior to 1961. In this work we analyse all the data from the third digitisation (2581 images) and use the data from the second digitisation for the missing dates (5943 images)\footnote{The analysis presented by \citet{chatzistergos_analysis_2017} and \citet{chatzistergos_exploiting_2016,chatzistergos_analysis_2018} was done with the data from the first 8-bit digitisation and the second digitisation for periods after and before 02/03/1960, respectively.}. 
The full-disc Ko data have been digitised two times, once with 8-bit depth \citep{makarov_22-years_2004} and once with 16-bit depth \citep{priyal_long_2014}. Here we use the Ko data from the 8-bit digitisation. 
The MW data have been digitised two times, once with 8-bit depth \citep{foukal_behavior_1996} and once with 16-bit depth \citep{lefebvre_solar_2005}. Here we use the 16-bit MW data. The Me data have been partially digitised at different periods with various set-ups. Some images have been digitised multiple times. In these cases we kept only those from the most recent digitisation.
The Sc, and WS data have only partially been digitised. Merely 18 and 452 plates out of 3131 and 4824 documented existing plates from the Sc and WS archives,  respectively, were digitised. 
Most of the digitised images were stored in FITS files, the Mi data from the 3rd digitisation and most of the Sc and WS data were stored in TIFF files, while the MM and the remaining Sc and WS data were stored in JPG compressed files. The various digitisations of the Me data were stored in GIF, TIFF, JPG, and FITS file formats. In a first step of our analysis all of the GIF, TIFF, and JPG files were converted to FITS files.

Calibration of the CCD employed for the digitisation was applied on the data from Ar observatory. MM, Sc, WS, and Mi data from the 3rd digitisation lack this processing step, while for Mi data from the earlier digitisations, as well as the Me, Ko, and MW data it is unclear whether this calibration was applied or not. 

Figure \ref{fig:processedimagessamedayoriginal} provides examples of images from each historical dataset taken on roughly the same five days, with the exception of panel Me1. The images were selected such that an observation from Sc or WS exists within an interval of 2 days. 
This period is unfortunately restricted to 1959--1969. Furthermore, panel Me1) shows the oldest available Ca II K observation from the analysed archives taken on 17/11/1893.
The brightness scale is individually normalized to cover the values from the minimum (black) to the maximum (white) within the solar disc in each image. The figure shows, for instance different brightness properties, saturated regions in the Ar images (Ar1, Ar3, and Ar5), overexposed plage regions in MM observations (MM1 and MM5),  or very bright small-scale artefacts in Mi observations (Mi1). 
See \citet{chatzistergos_analysis_2017} for a further discussion of problems affecting such data. 
Figure \ref{fig:processedimagessamedayoriginal} is discussed in more detail in the Appendix \ref{sec:sorting}.

\begin{table*}
	\caption{List of Ca~II~K archives analysed in this study.}
	\label{tab:observatories}     
	\centering                      
	\begin{tabular}{l*{9}{c}}       
		\hline\hline                
		Observatory & Abbreviation&Years & \multicolumn{2}{c}{Images} &SW  & Disc size& Data type & Pixel scale  & Ref.\\
		&  & &Total &Used & [$\text{\AA}$] &  [mm]& [bit]& [$"/$pixel]& \\
		\hline                                  
		Arcetri					   &Ar		 & 1931--1974				  &5133 &4825 &0.3 &$ 65$   &16   			    &2.4 	  			      & 1\\
		Kodaikanal\tablefootmark{a}&Ko		 & 1907--1999				  &22158&19291&0.5 &$ 60$   &8    			    &1.3 	  				  & 2\\ 
		McMath-Hulbert			   &MM		 & 1948--1979				  &5912 &4911 &0.1 &-	    &8   			    &3.5         			  & 3\\
		Meudon	   				   &Me		 & 1893--2018\tablefootmark{b}&20986&17605&0.15&$ 65$   &16					&0.9--11\tablefootmark{c} & 4\\
		Mitaka  				   &Mi		 & 1917--1974				  &8585 &4193 &0.5 &variable&16\tablefootmark{d}&0.9, 0.7\tablefootmark{e}& 5\\
		Mount Wilson 			   &MW		 & 1915--1985				  &36340&31430&0.2 &$ 50$   &16  			    &2.7	  				  & 6\\
		Rome 					   &Rome/PSPT& 1996--2018				  &3292 &3292 &2.5 &$\sim27$&16			        &2\tablefootmark{f}	  	  & 7\\
		Schauinsland			   &Sc 		 & 1958--1965\tablefootmark{g}&18   &18   &-   &variable&16  			    &1.7, 2.6\tablefootmark{h}& 8\\
		Wendelstein  			   &WS 		 & 1947--1977\tablefootmark{g}&452  &407  &-   &variable&16   			    &1.7, 2.6\tablefootmark{h}& 8\\
		\hline
	\end{tabular}
	\tablefoot{Columns are: name and abbreviation of the observatory, period covered, the number of available and used observations, spectral width of the spectrograph/filter, disc size in the photographic observation, digitisation depth, pixel scale of the digital image, and the bibliography entry. 		
		\tablefoottext{a}{This archive has been recently re-digitised with 16-bit data type extending the dataset from 1904 to 2007 \citep{priyal_long_2014,chatterjee_butterfly_2016}, but here we use the 8-bit data from a former digitisation.}
		\tablefoottext{b}{The observations after 13/05/2002 are recorded with a CCD camera, while data are stored on photographic plates until 27/09/2002.} 
				\tablefoottext{c}{The highly variable resolution is due to different set-ups employed for the digitisation. The average value for the photographic data (1893--2002) is 2.3$"/$pixel, while for the CCD data (2002--2017) is 1.5$"/$pixel.} 
		\tablefoottext{d}{All these data are with 16-bit digitisation depth, however they were the result of 2 different digitisation runs. The data after 02/03/1960 as well as the first 10 observations of the earlier years were recently re-digitised with a different scanning instrument.}
		\tablefoottext{e}{The two values correspond to the data from the second and third digitisation.}
		\tablefoottext{f}{The pixel scale is for the resized images.}\tablefoottext{g}{These are the years covered by the digitised observations, the WS archive goes back to 1943, while the Sc covers the period 1944--1966.}\tablefoottext{h}{The two values correspond to the JPG and TIFF files, respectively.}
	Links to data. Ar: \url{https://venus.oa-roma.inaf.it/solare/cvs/arcetri.html}, Ko: \url{https://kso.iiap.res.in/new/archive/input}, MM: \url{ftp://ftp.ngdc.noaa.gov/STP/space-weather/solar-data/solar-imagery/chromosphere/calcium/mcmath/}, Me: \url{http://bass2000.obspm.fr/search.php}, Mi: \url{http://solarwww.mtk.nao.ac.jp/en/db_ca.html#fits}, and Rome/PSPT: \url{https://venus.oa-roma.inaf.it/solare/}}
		\tablebib{
		(1)~\citet{ermolli_digitized_2009}; (2) \citet{makarov_22-years_2004}; (3) \citet{mohler_mcmath-hulbert_1968}; (4) \citet{mein_spectroheliograms_1990}; (5) \citet{hanaoka_long-term_2013}; (6) \citet{lefebvre_solar_2005}; (7) \citet{ermolli_photometric_2007}; (8) \citet{wohl_old_2005}.}
\end{table*}

The modern observations (i.e. those taken with a CCD detector) analysed in this study were acquired with at the Meudon and the Rome observatories.
Meudon observatory installed a CCD camera on 13/05/2002 and continued observations with the same instrumental set-up, while storing data also on photographic plates up to 27/09/2002.
The Rome observations were acquired with the Rome/PSPT, which is a 15 cm, low-scattered-light, refracting telescope designed  for synoptic  solar observations characterized
by  0.1\% pixel-to-pixel 
relative photometric precision
\citep{coulter_rise/pspt_1994}. %
It is operated at  
Monte Porzio 
Catone Observatory by the INAF Osservatorio Astronomico di Roma  \citep{ermolli_prototype_1998,ermolli_photometric_2007}. The Rome/PSPT has been acquiring daily full-disc observations since May 1996, but only from mid 1997 onwards with final instrumental set-up. 
The data are taken daily on 2048$\times 2048$ CCD arrays with various narrow-band  
interference filters within a few minutes from each other. In this study, we use observations carried out with filters centred on the 
Ca~II~K line after the standard instrumental calibration has been applied \citep[][]{ermolli_measure_2003}. These images are the sums of 25 frames.

\begin{figure*}[t]
	\centering
	\includegraphics[width=1\linewidth]{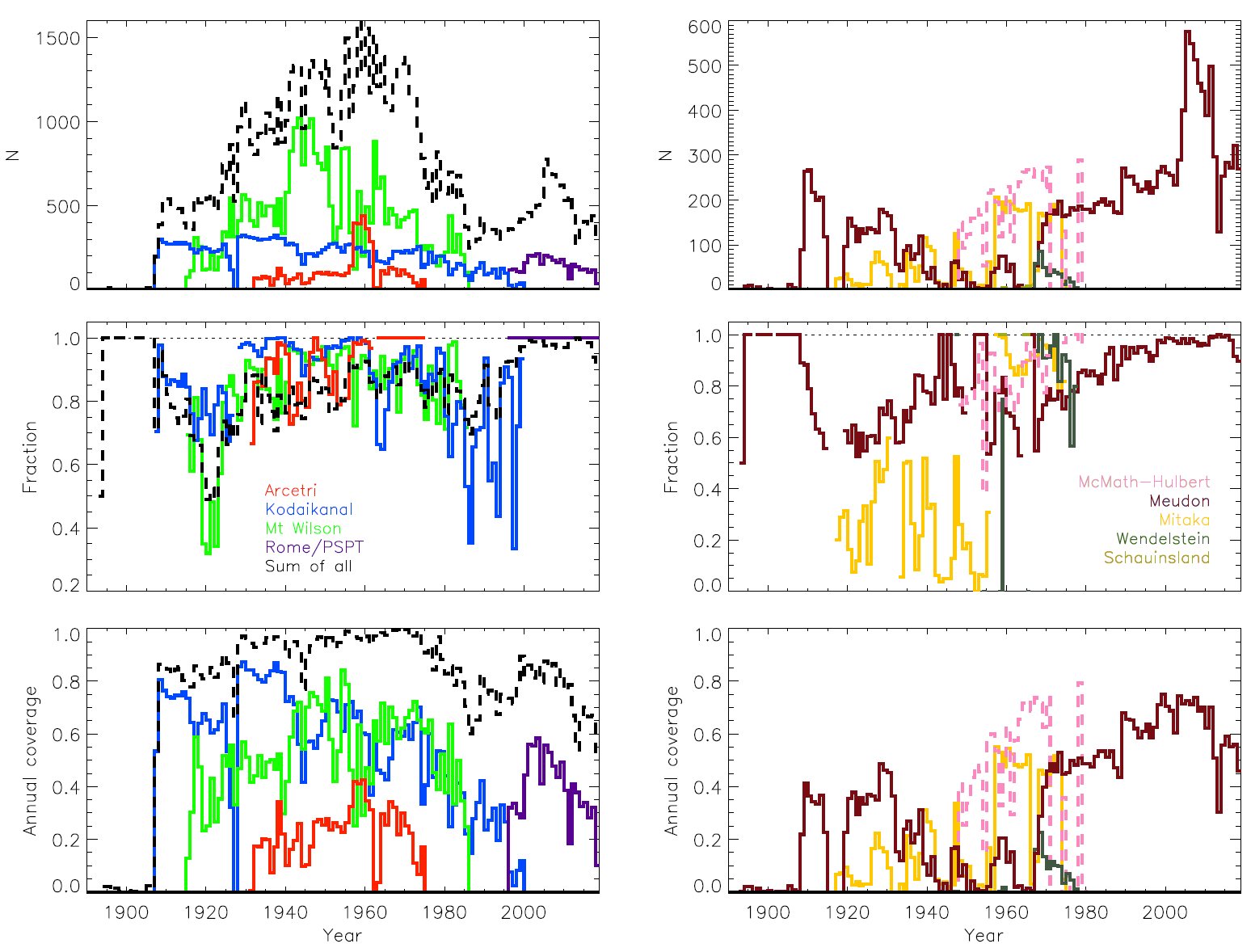}
	\caption{The number of images per year from the different archives (see legend) considered in this study (top row), the fraction of the available images from each archive per year that we used for our final analysis (middle row), and the fraction of days per year covered by each archive (bottom row). For the sake of clarity, the analysed series were divided in two groups, shown in the left and right columns. Also shown is the sum from all archives (black line in the left panels). }
	\label{fig:ndata_fraction}
\end{figure*}

Figure \ref{fig:ndata_fraction} shows the annual distribution of the number and the fraction of the images from all the archives we used for our analysis.
The fraction of days per year covered by each archive is also shown in Fig. \ref{fig:ndata_fraction}, while Table \ref{tab:commondaysobservatories} lists the number of the common observing days for the various pairs of all the archives used here.

As can be seen in Fig. \ref{fig:ndata_fraction}, Me is the longest series among the ones included in this study, with data extending from 1893 to 2018. Me includes observations over 14512 days out of which 10880 days are from the photographic data. However, there are currently very few or no digitised data  over the periods 1893--1906, 1914--1918, and 1940--1961. Ko is the second longest series, with data for 18963 days in total with about one observation per day. However the amount of Ko images per year decreases with time and especially after 1990 remains very low. 
MW is the third longest series studied here. It includes observations from 13491 days, and generally multiple images were taken for each day of observation.
We notice that the amount of MW data during 1940--1950 is roughly 2--3 times larger than during other periods.
With 4901 days over a period of 32 years, MM is the shortest series employed here other than Sc, and WS which are incomplete. 
Ar and Mi have less observations per year than the other archives, thus including 3573 and 3987 days of observation, respectively.
Ar data exhibit a peak of image availability between 1957 and 1962.
Due to the limited amount of Sc and WS data and taking into account the similarity of these observations, which were part of the same observing program of the Kiepenheuer-Institut f\"ur Sonnenphysik, we merged these two archives and we report results from both of these datasets as a single series.

\subsection{Synthetic data}
\label{sec:syntheticdata}
We also employed the synthetic images created by \cite{chatzistergos_analysis_2018}.
In particular, in this study we used the subsets 1, 6, and 8 from \cite{chatzistergos_analysis_2018} in their unchanged form. 
The basis for the synthetic data were the Rome/PSPT images covering the period 2000--2014. Briefly:
\begin{itemize}
	\item  For subset 1, we employed contrast images, imposed on them the centre-to-limb variation (CLV, hereafter) computed as the average over all the available Rome/PSPT images, and applied a linear characteristic curve (CC, hereafter). We did not introduce any other image problems so to create images that describe the best quality historical data. Subset 1 consists of 500 observations equally spaced in time during the period 2000--2014.
\item Subset 6 was created by using 10 observations, on which we applied the same CLV and CC as in subset 1. However, prior to the application of the CC, we imposed 20 different inhomogeneity patterns on the images. Each of these patterns was derived to replicate some of the common problems in the historical data and was used with 10 severity levels. Thus subset 6 includes 2000 images representing observations with quality varying from very good to very poor.
\item Subset 8 includes 2000 images, each one created with a random set of problems and severity level. In this sample, the CC is in the form of a 3rd degree polynomial with random parameters for each image. The CLV pattern, the inhomogeneities with the corresponding severity levels, and the vignetting were chosen randomly to replicate the most common features of the historical series.
\end{itemize}

\subsection{Time-series of plage areas}

We will also compare our results with other existing plage area time-series.
We considered the series of plage areas derived by
\cite{kuriyan_solar_1982},  \cite{foukal_behavior_1996}\footnotemark, \cite{ermolli_digitized_2009}, \cite{ermolli_comparison_2009}, \citet{tlatov_new_2009}, \cite{bertello_mount_2010}, \cite{chatterjee_butterfly_2016}, \cite{priyal_long-term_2017}, \cite{singh_variations_2018}, and those presented in the National oceanic and atmospheric administration Solar Geophysical data reports\footnotemark[\value{footnote}] (SGD, hereafter). 
\footnotetext{Available at \url{https://www.ngdc.noaa.gov/stp/solar/calciumplages.html}}

In brief, SGD provided monthly tables\footnote{The report books are available at \url{https://www.ngdc.noaa.gov/stp/solar/sgd.html}. More information on the derivation of the plage areas is given in report No. 515-supplement from July 1987.} for the plage areas derived from the MM (06/1942--09/1979), MW (10/1979--09/1981), and Big Bear solar observatory (10/1981--11/1987). The plage areas from the MM and MW archives were obtained from the physical photographs and drawings. According to \citet{foukal_curious_1993}, this was done manually with prints of overlay surfaces, which included circles of different sizes and different positions on the disc in order to take the foreshortening into account. Sunspots were included in the plage areas. The processing of the Big Bear solar observatory data is described in \cite{marquette_calcium_1992}.
Similarly, \cite{kuriyan_solar_1982} derived plage areas from the physical Kodaikanal photographs.
\cite{foukal_behavior_1996} used uncalibrated MW data from the 8-bit digitisation and produced the time-series (termed $A_\mathrm{p}$ index) by manually selecting the plage regions in every image. 
\cite{ermolli_digitized_2009} used the available wedge exposures on the Ar plates to construct the CC and to calibrate the Ar images. 
\cite{ermolli_comparison_2009} calibrated the Ar, Ko (8-bit), and MW (16-bit) data with a linear relation, derived for each image separately, and segmented the images with constant thresholds for the contrast and feature size. 
\citet{tlatov_new_2009} processed the data from Ko (8-bit), MW (16-bit), and Sacramento Peak observatories. The data were linearly scaled so that their CLV matched a reference one \citep[see][]{chatzistergos_analysis_2017,chatzistergos_analysis_2018} and then segmented with a threshold of a constant multiplicative factor to the full width at half-maximum of the distribution of intensity values of the image (including the CLV).
\citet{bertello_mount_2010} processed uncalibrated MW 16-bit data. 
They calculated the histogram of contrast image values and defined a parameter related to plage areas. This index does not carry information on the absolute plage area, and needs to be calibrated to a reference plage area series independently; see Sect. \ref{sec:bertechat} for more details.
\cite{chatterjee_butterfly_2016} used uncalibrated 16-bit Ko data and identified the plage regions as those with $C>C_{\mathrm{median}}+\sigma$, where $C$ is the contrast of each pixel, $C_{\mathrm{median}}$ the median contrast of the disc and $\sigma$ the standard deviation of contrasts.
\cite{priyal_long-term_2017} calibrated the 16-bit Ko data by applying an average relation derived from the data that include a calibration wedge and then segmented the data with constant thresholds to derive a plage area series. \cite{singh_variations_2018} analysed the 16-bit Ko data. They manually selected the observations that were used for their plage areas time-series with the purpose of excluding data that have potentially been taken with different bandwidths or were centred outside the line core. It is worth noting that no data after 1976 were used in the time series of \cite{singh_variations_2018}. 

We also considered one plage area record derived from Ca~II~K filtergrams taken at the San Fernando observatory (SFO) with the CFDT1 and CFDT2 (Cartesian full-disc telescopes 1 and 2, respectively). 
CFDT1 has been operating since mid-1986, producing images of 512 x 512 pixels with a pixel size of 5.12$''/$pixel. CFDT2 started operation in mid-1992 producing images of 1024 x 1024 pixels at a pixel size of 2.5$''/$pixel.
Both telescopes use filters centred on the Ca~II~K line with bandwidth of 9 \AA~ \citep{chapman_solar_1997}, i.e. considerably broader than the historical observations and hence with stronger photospheric contributions.

It is worth noting that the above series were derived with rather different processing techniques and from different data, also obtained from diverse digitisations of the same archive, or from various samples of observations therein. 
Furthermore, none of the above studies attempted to combine plage series from different image sources.

\section{Image processing}
\label{sec:preprocessing}
All together, we have analysed 99584 images from 8 different historical archives, of which only 82680 were used for our analysis. We also used 3292 (all available) images from the modern dataset of Rome/PSPT. We ignored all pathological cases . Specifically, we used 4825 Ar, 19291 Ko, 4911 MM, 17605 Me, 4193 Mi, 31430 MW, 3292 Rome/PSPT, 18 Sc, and 407 WS images.
This required development of automated and robust correction and calibration methods the main steps of  which are outlined in the following. Appendices \ref{sec:sorting} and \ref{sec:preprocessingstep} describe in more detail how we sorted  and pre-processed the data, as well the characteristics of the individual archives.

\begin{figure*}[t!]
	\centering
	\includegraphics[width=1\linewidth]{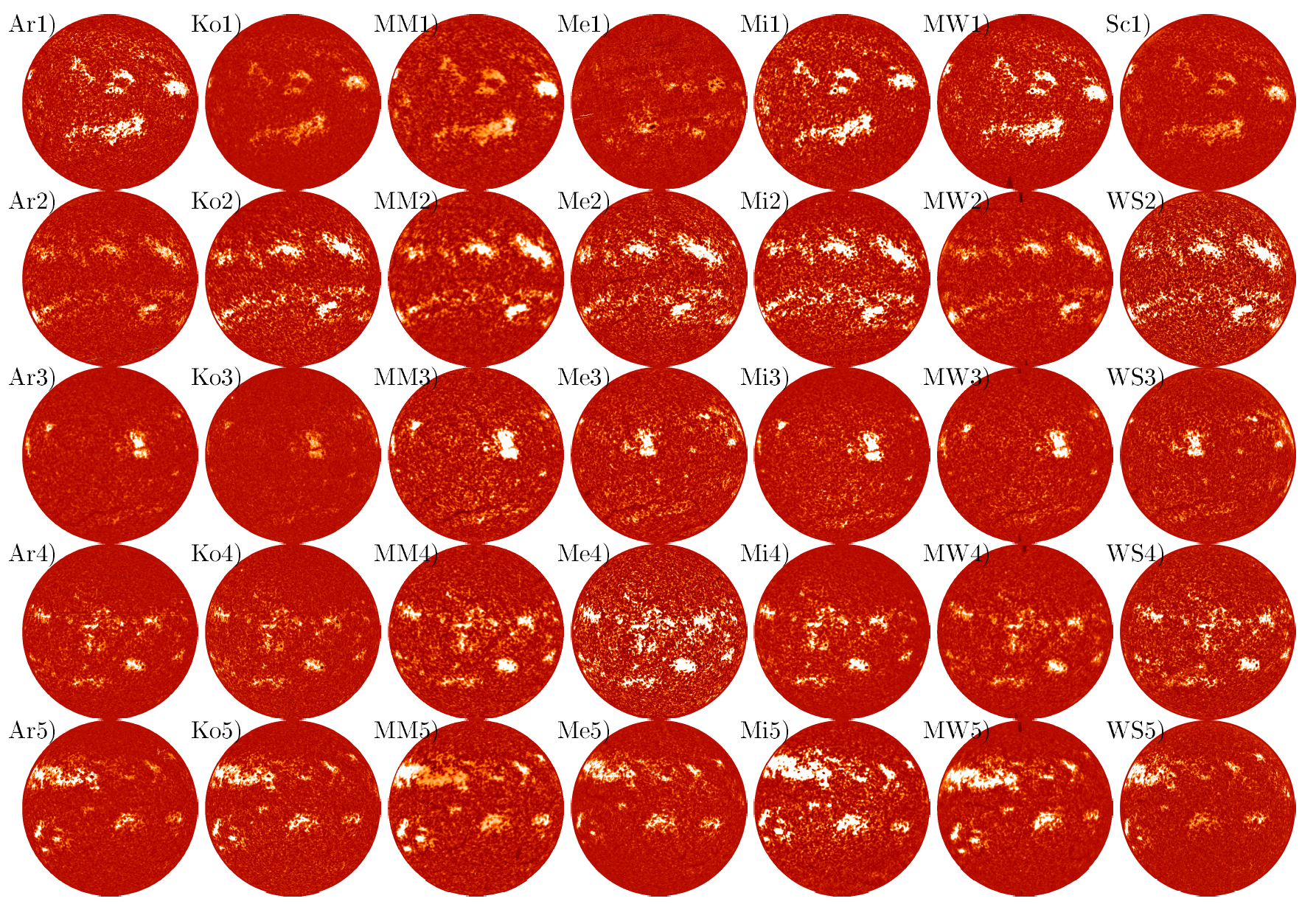}
	\caption{Calibrated and CLV-compensated images derived from the observations shown in Fig. \ref{fig:processedimagessamedayoriginal}. The images have been corrected for artefacts as described in Sect. \ref{sec:preprocessingstep} and are compensated for ephemeris. All images are saturated in the range [-0.5,0.5].}
	\label{fig:processedimagessamedayflat} \end{figure*}

\begin{figure*}[t]
	\centering
	\includegraphics[width=1\linewidth]{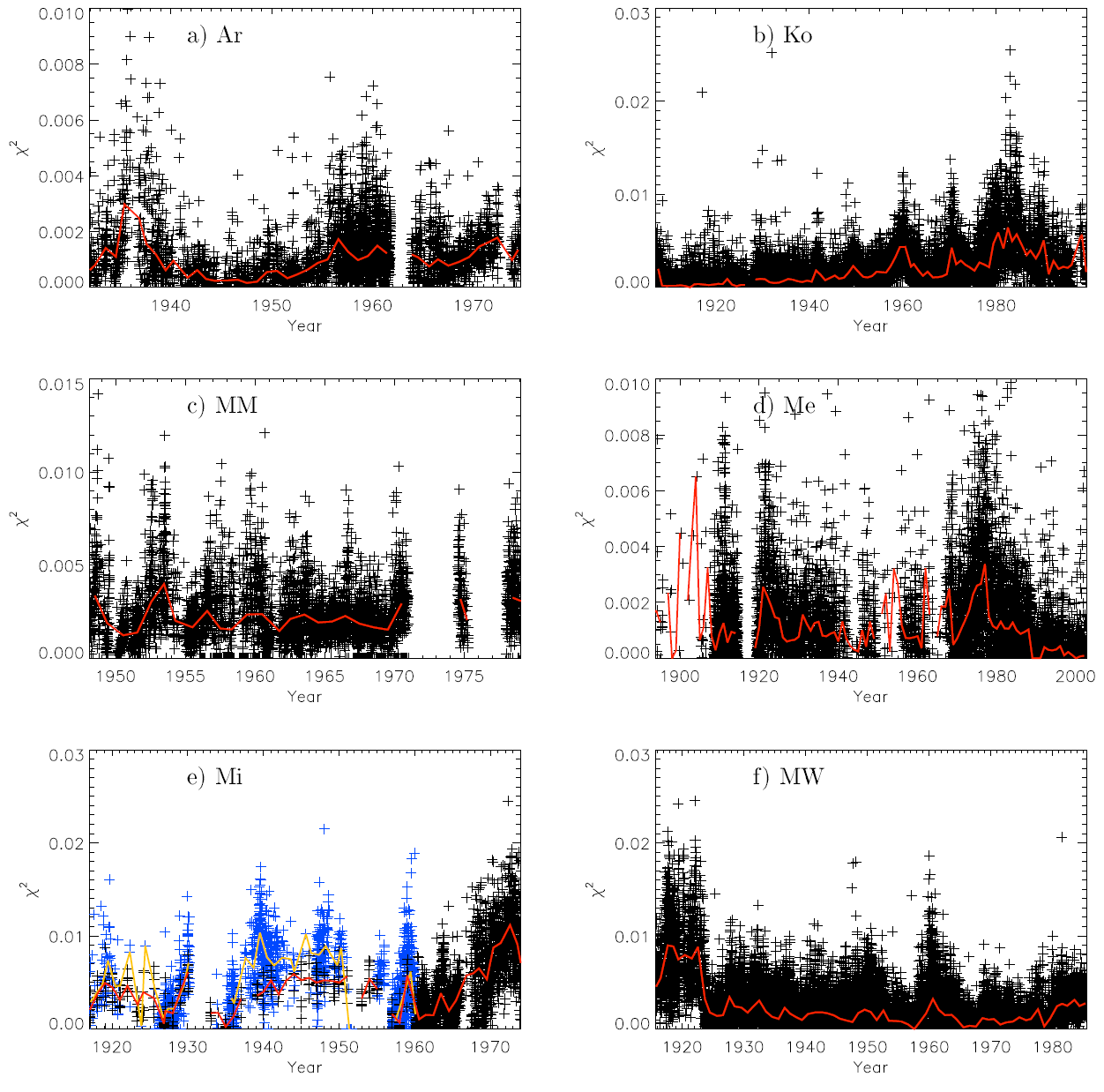}
	\caption{Temporal variation of the reduced $\chi^2$ of the fit performed on the QS CLV to determine the CC during the photometric calibration of the images. The values shown here are for all images from the Ar (a), Ko (b), MM (c), Me (d), Mi (e), and MW (f) archives. For Mi we show the values for the images of the earlier (blue) and the more recent (black) digitisations separately. The red solid (yellow for the earlier Mi digitisation) line shows annual median values. }
	\label{fig:chi2vstime}
\end{figure*}

\begin{figure*}
	\centering
	\includegraphics[width=1\linewidth]{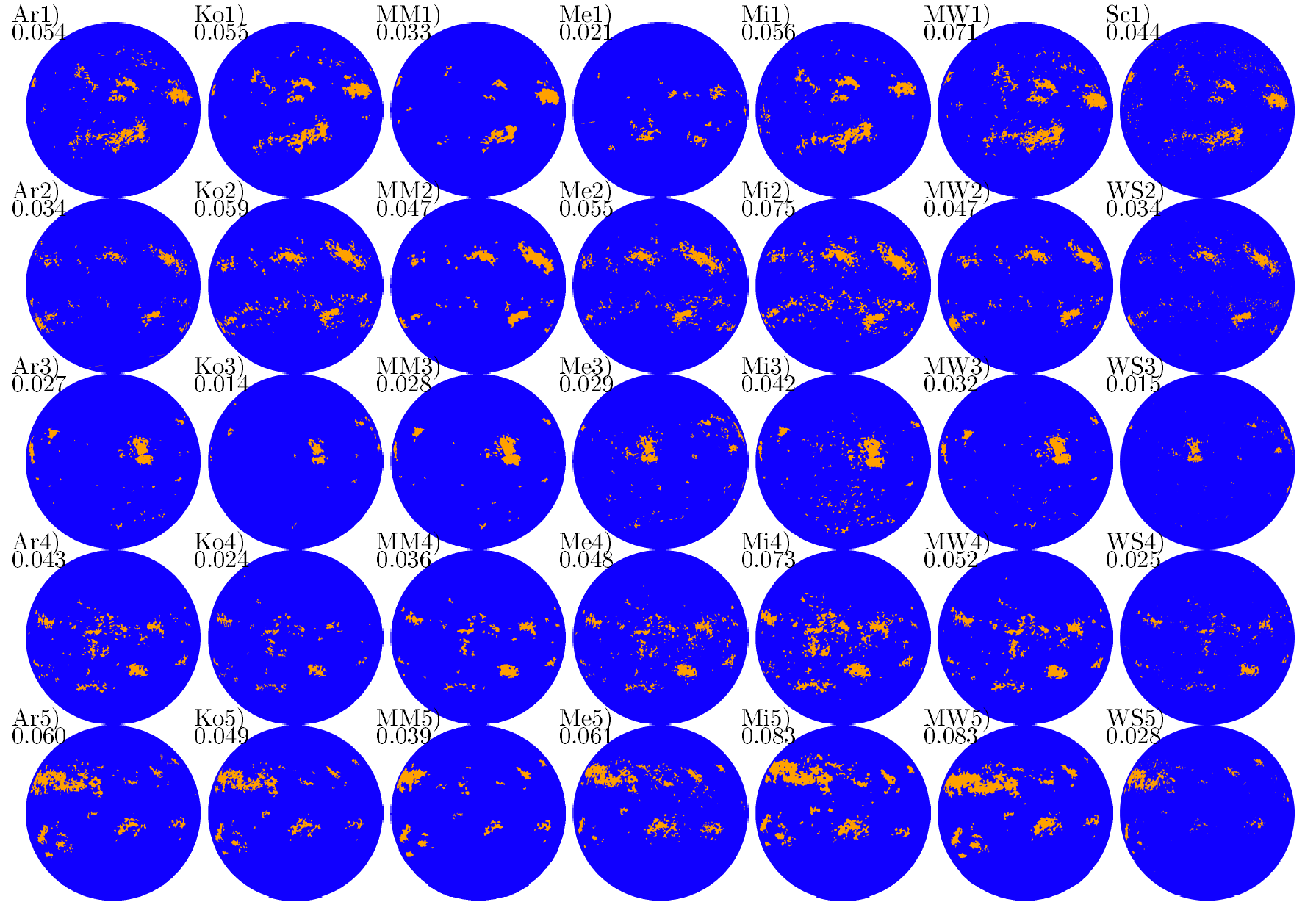}
\caption{Segmentation masks derived from the observations shown in  Fig. \ref{fig:processedimagessamedayoriginal}. Disc features shown are plage (orange) and quiet Sun (blue). The disc fraction covered by plage is given on the upper left side of each mask.}
	\label{fig:processedimagessamedaymask} \end{figure*}

\begin{figure*}[t]
	\centering
		\includegraphics[width=1.0\linewidth]{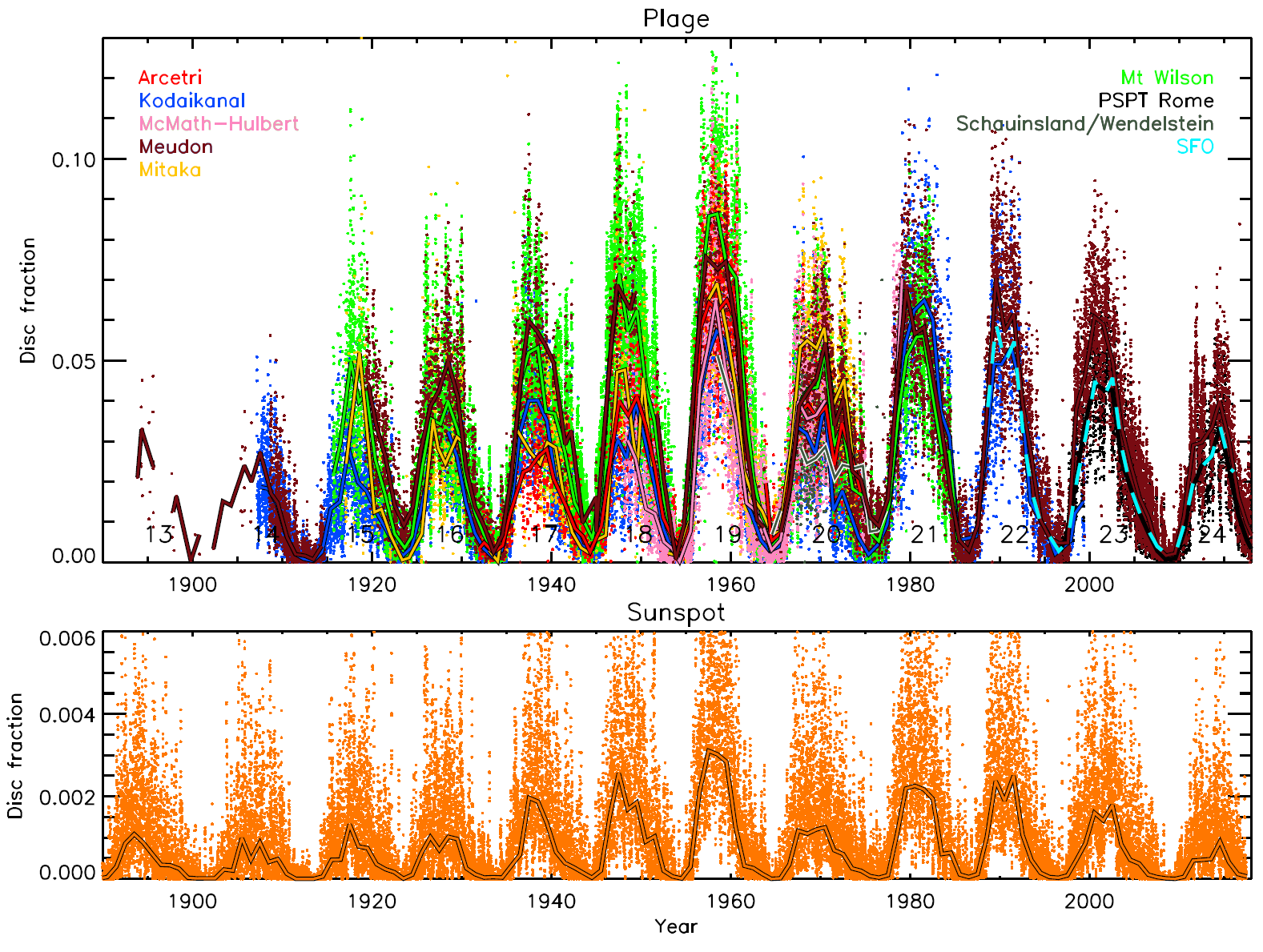}
	\caption{Top panel: Fractional disc coverage by plage as a function of time, derived with the NR thresholding scheme applied with the same parameters on images from the Ar (red); Ko (blue); MM (pink); Me (brown); Mi (orange); MW (green); Sc/WS (dark green); Rome/PSPT (black) archives. Also shown are the plage areas from SFO (light blue). The numbers under the curves denote the conventional SC numbering. Bottom panel: Sunspot areas  from \citet[][]{balmaceda_homogeneous_2009}. Individual small dots represent daily values, while the thick lines indicate annual median values.}
	\label{fig:discfraction_time}
\end{figure*}

\begin{figure*}[t]
	\centering
			\includegraphics[width=0.75\linewidth]{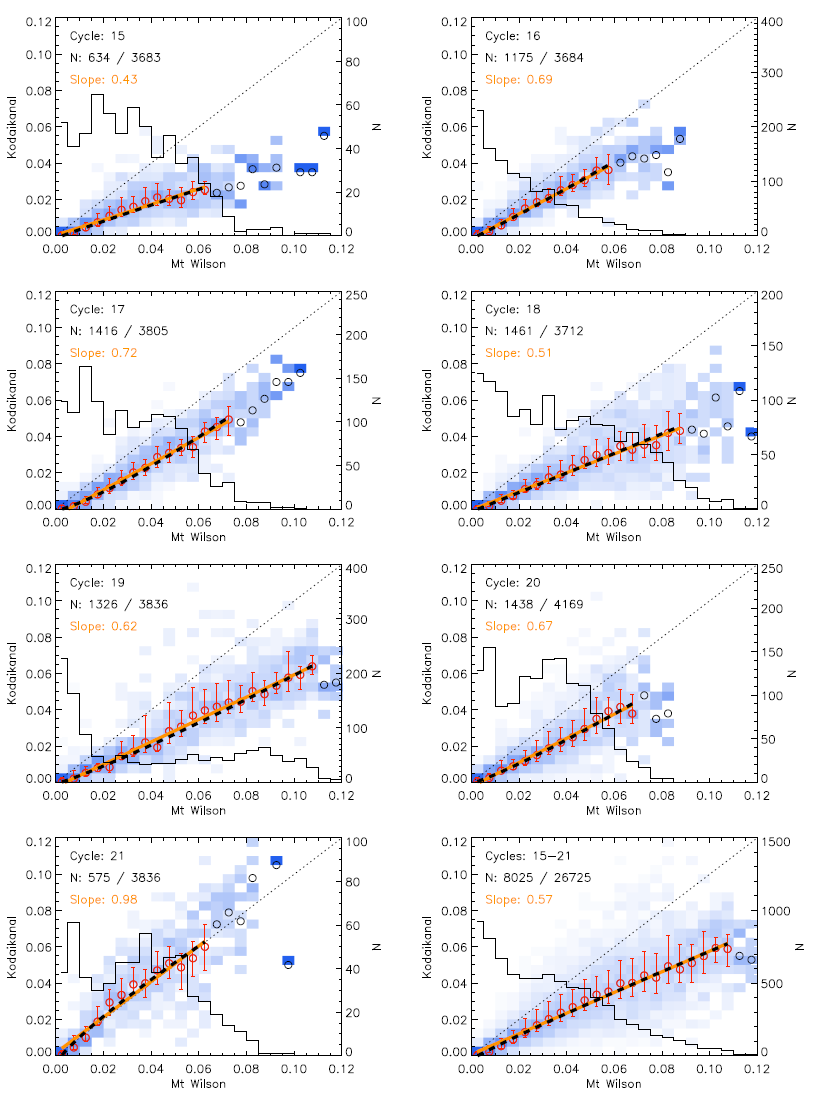}
	\caption{Probability distribution functions (PDF) for fractional disc coverage by plage derived from Ko data, within area bins of 0.005, as a function of the disc fractions from MW for the same days. The PDF are colour-coded with white being 0 and bright blue 1. The Ko series is taken as the reference for this particular representation. Circles denote the average value within each column. For columns with less than 20 days of data these circles are shown in black, otherwise in red. We also show the asymmetric 1$\sigma$ interval for each column. Two different fits to the average values are over-plotted; a linear fit (yellow) and a power law fit (black). The number of days included in each column is over-plotted with a solid black line (see right-hand axis). The dotted black line has a slope of unity and represents the expectation value. Each panel shows the PDF matrix for a given SC, while the lower right panel is for the whole interval of overlap between the compared series. The number of overlapping days used to construct each matrix (N) and the total number of days within each time interval is written in the plots. The slope of the linear fit is also given.}
	\label{fig:komwcycles14-25}
\end{figure*}

\begin{figure*}[t]
	\centering
			\includegraphics[width=0.8\linewidth]{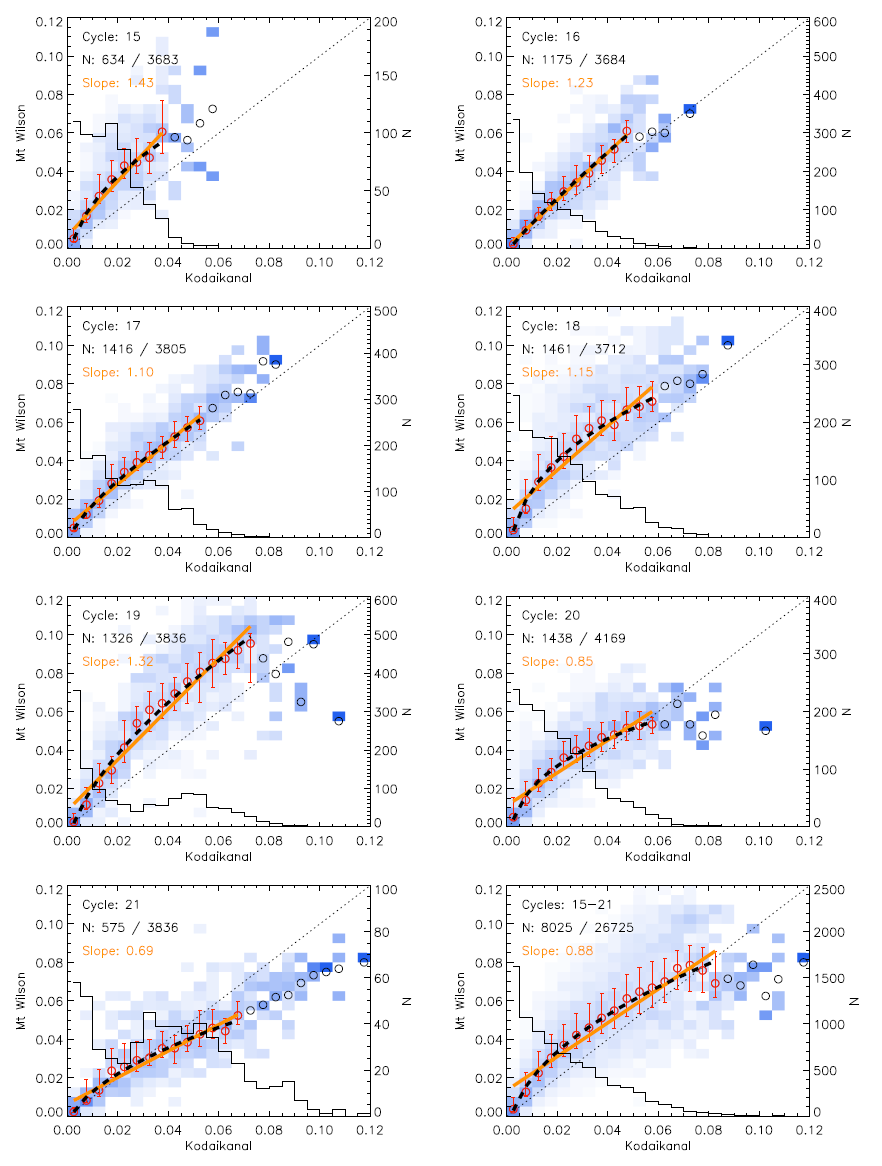}
	\caption{Same as Fig. \ref{fig:komwcycles14-25} but using MW plage area series as the reference and Ko as the secondary set.}
	\label{fig:mwkocycles14-25}
\end{figure*}

\begin{figure*}[t]
	\centering
			\includegraphics[width=0.8\linewidth]{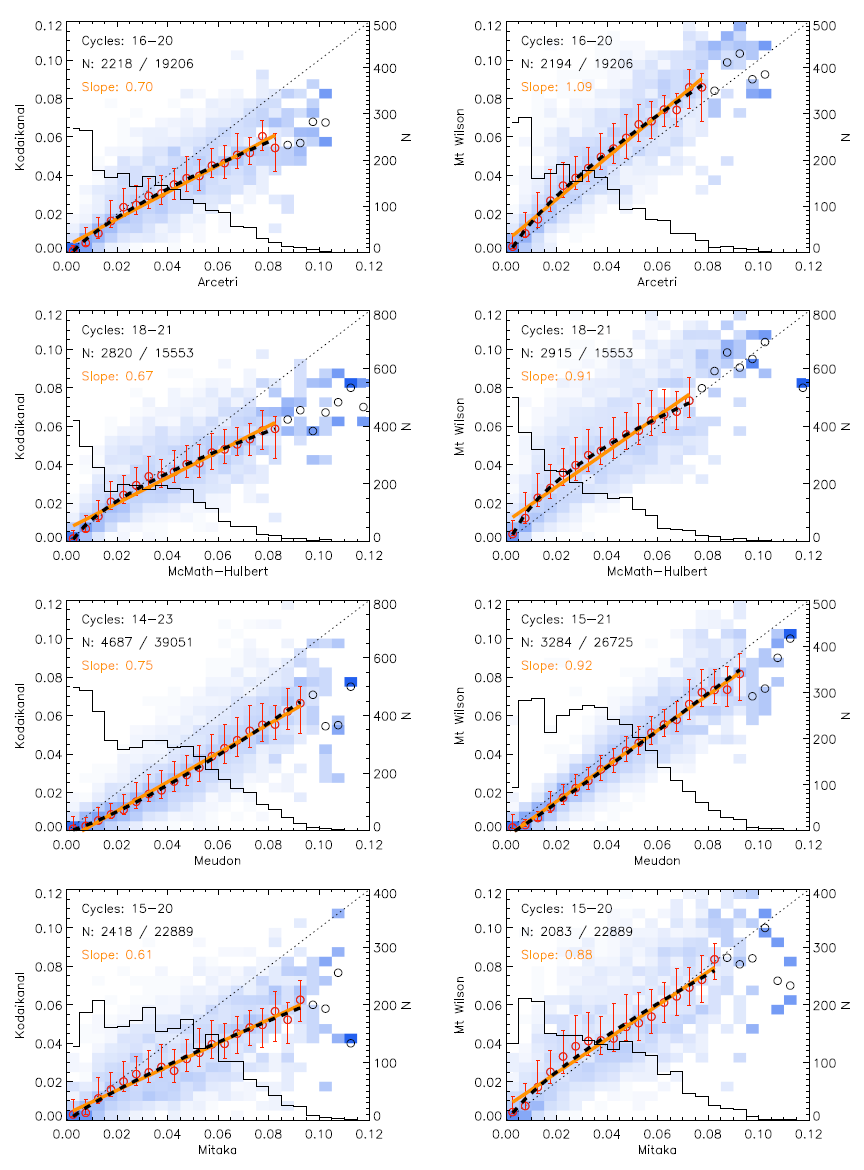}
	\caption{Same as Fig. \ref{fig:komwcycles14-25} but with Ko (left) and MW (right) plage area series as reference and Ar (1st row), MM (2nd row), Me (3rd row), and Mi (4th row) series as the secondary, by considering the whole period of observations over which the respective datasets overlap (see Table \ref{tab:commondaysobservatories}).}
	\label{fig:komwothers}
\end{figure*}

\begin{figure*}[t]
	\centering
			\includegraphics[width=0.8\linewidth]{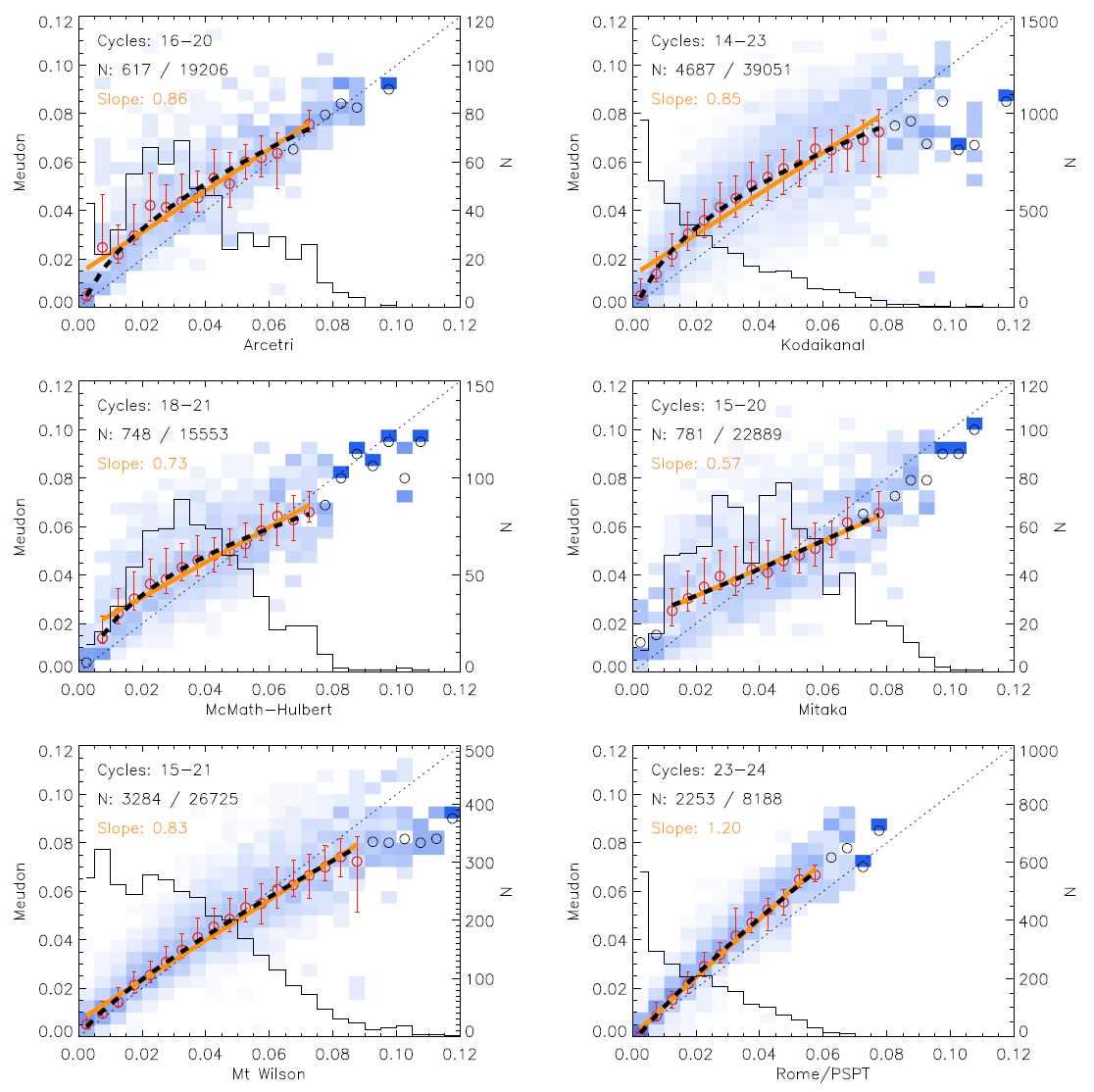}
	\caption{Same as Fig. \ref{fig:komwcycles14-25} but with Me plage area series as reference and Ar, Ko, MM, Mi, MW, and Rome/PSPT series as the secondary, by considering the whole period of observations over which the respective datasets overlap (see Table \ref{tab:commondaysobservatories}).}
	\label{fig:meothers}
\end{figure*}

\begin{figure*}[t]
	\centering
			\includegraphics[width=0.9\linewidth]{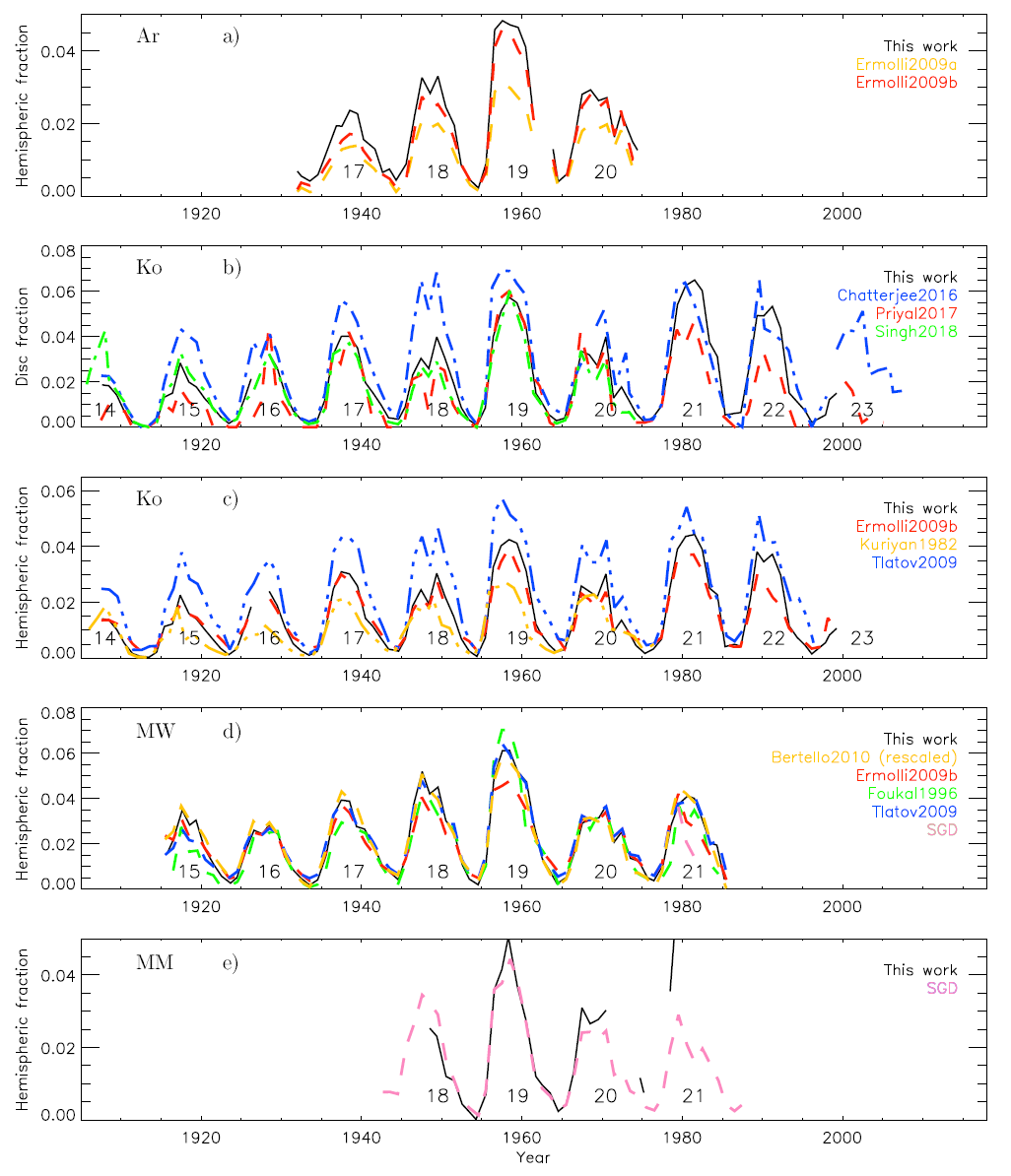}
	\caption{Comparison of the plage areas derived here (black) to other published series based on the same datasets: (a)) \citet[][yellow]{ermolli_digitized_2009} and \citet[][red]{ermolli_comparison_2009} for Ar data; (b)) \citet[][blue]{chatterjee_butterfly_2016}, \citet[][red]{priyal_long-term_2017}, and \citet[][green]{singh_variations_2018} for Ko areas in disc fractions;  (c)) \citet[][red]{ermolli_comparison_2009}, \citet[][yellow]{kuriyan_solar_1982}, and \citet[][blue]{tlatov_new_2009} for Ko areas in fractions of a hemisphere; (d)) \citet[][yellow]{bertello_mount_2010}, \citet[][red]{ermolli_comparison_2009}, \citet[][green]{foukal_behavior_1996}, \citet[][blue]{tlatov_new_2009}, and SGD (pink) for MW data; (e)) SGD (pink) for MM data. The curves are annual median values. All panels except b) are in fractions of hemisphere corrected for projection effects. The numbers under the curves denote the conventional SC numbering.}
	\label{fig:discfraction_plage}
\end{figure*}

\begin{table*}
	\caption{Number of common days for various pairs of the 9 Ca~II~K archives analysed in this study.}
	\label{tab:commondaysobservatories}
	\centering
	\begin{tabular}{l*{8}{c}}
		\hline\hline
		\small
&Ar& Ko& Me& MM& MW& Mi& Rome/PSPT& Sc/WS\\
		\hline
Ar& 3573& 2218& 617& 1308& 2194& 839& 0 & 109\\
Ko& 2218& 18963& 4687& 2820& 8025& 2418& 33& 242\\
Me& 617& 4687& 14512& 748& 3284& 781& 2253& 159\\
MM& 1308& 2820& 748& 4901& 2915& 1268& 0 & 172\\
MW& 2194& 8025& 3284& 2915& 13491& 2083& 0 & 256\\
Mi& 839& 2418& 781& 1268& 2083& 3987& 0 & 127\\
Rome/PSPT& 0 & 33& 2253& 0 & 0 & 0 & 3287& 0 \\
Sc/WS& 109& 242& 159& 172& 256& 127& 0 & 411\\
		\hline
	\end{tabular}
\end{table*}

\subsection{Photographic calibration and CLV compensation}
\label{sec:processing_calibration}
After the pre-processing, the images analysed in our study were photometrically calibrated and compensated for the CLV as described by \cite{chatzistergos_analysis_2018}. 
Briefly, it is an iterative process to identify the quiet Sun (QS, hereafter) background by applying a running window median filter after the active regions have been singled out from the solar disc. 
Assuming that the QS does not significantly vary with time \citep[as suggested by several studies in the literature, see][for more details]{chatzistergos_analysis_2018}, we relate the derived density QS CLV to the logarithm of intensity QS CLV we get from modern CCD-based observations. We thus derive a linear CC between the density and logarithm of intensity, which is used to calibrate the image and the QS CLV.
We then produce contrast images as $C_i=(I_i-I^{\mathrm{QS}}_i)/I^{\mathrm{QS}}_i$, where $C_i$, $I_i$ is the contrast and the intensity of the $i$th pixel, respectively, while $I^{\mathrm{QS}}_i$ is the level of the QS intensity in the $i$th pixel.
For our analysis, we only considered pixels within $0.99R$ to avoid the influence of the uncertainties in the radius estimates. This corresponds to $\mu=\cos{\theta}=0.14$, where $\theta$ is the heliocentric angle. The accuracy in determining the edge of the disc is considerably higher for the Rome/PSPT images, so for these images we considered pixels within $0.9999R$.
Figure \ref{fig:processedimagessamedayflat} shows the resulting calibrated and CLV-compensated images for the observations displayed in Fig. \ref{fig:processedimagessamedayoriginal}.
The Me and Rome/PSPT CCD data calibrated for instrumental effects were processed to compensate for the CLV in the same way as the historical data.

We consistently applied the image processing described above to data from different archives.
In Fig. \ref{fig:chi2vstime} we show the temporal variation of the reduced $\chi^2$ of the fits to derive the CC for the data from Ar, Ko, MM, Me, Mi, and MW observatories. The goodness of the fit in this step mainly depends on three factors: the presence of strong artefacts over the recorded image of the solar disc, exposure problems, and the similarity of the QS CLV measured on the historical data to the one we use as a reference. The latter is essentially giving us an indication for changes in the bandwidth or the central wavelength of the observation. 
Hence abrupt and systematic changes in the determined 	$\chi^2$ of the fits are indications for changes in the instruments or in the observing conditions. 
It should be noted, however, that this is not a very strict criterion to identify all possible quality transitions in the archives, since multiple effects can counterbalance each other (e.g. use of narrower bandwidth when the observations suffer from vignetting effects). 
Figure \ref{fig:chi2vstime} shows that the values of the plotted quantity do not significantly vary with time for Ar and MM, while they gradually increase with time for Ko. 
Besides, we notice a slight increase in the Ar values at approximately 1953, coinciding with the change in the spectrograph (see Appendix \ref{sec:sorting}). 
Furthermore, the values for MW show an abrupt jump before 08/1923. This is attributed to an instrumental change reported in the MW logbooks as the implementation of a different grating since 21/08/1923.
For the values of Mi we see a 
jump after 02/1966 which coincides with instrumental adjustments. 
We do not discern any difference in the $\chi^2$ of the fit around 1925, when the Mi observatory was relocated, however the data points are too few to derive any meaningful conclusions about this.
There are many changes in the $\chi^2$ of the fit for Me data, however we cannot derive any meaningful conclusions due to the low number of data for few years and the multiple digitisations within a given year. An exception is for the data around 1980, when there is a notable jump in the $\chi^2$ values. This can potentially be attributed to the digitisations, since the data in the period 1970--1980 belong to a different digitisation than the ones after 1980. Furthermore, considering only the period after 1980 when we have sufficient data which derive from the same digitisation, we note that there is a small decrease in the $\chi^2$ after 1990.

\subsection{Plage identification}
We performed the disc segmentation with a variant of the method presented in \citet[][NR, hereafter]{nesme-ribes_fractal_1996} as used e.g. by \citet{ermolli_comparison_2009}. 
This method assumes a Gaussian background brightness distribution, while magnetic features add a non-Gaussian contribution to the wings of the distribution: at the low and high contrast for dark spots and bright plage, respectively. We first compute the mean contrast, $\bar{C}$, and the standard deviation of the contrast, $\sigma _{C}$, over the disc. 
We identify pixels that have contrasts  within $\bar{C}\pm k\sigma_{C}$ for an array of values of $k$ (typically in the range 0.1 to 3.0). For these locations we calculate the mean contrast and the standard deviation.
The minimum of the calculated mean contrasts, $\bar{C}_{\mathrm{min}}$,  best represents the background QS regions. The idea is that network and plage 
skew the distribution and cause $\bar{C}$ to be shifted to positive values, while the darker spots and pores cover a sufficiently small part of the solar surface as not to dominate over the effect of plage and network. The value of $k$ that gives the lowest mean contrast is adopted as the best representation of the QS contrast. 
\citet{nesme-ribes_fractal_1996} found that the optimum threshold $K$ is the contrast that corresponds to a slightly higher value of $k$ than $k_{\bar{C}_{\mathrm{min}}}$. They considered values of $k$ between $0.3+k_{\bar{C}_{\mathrm{min}}}$ and  $0.6+k_{\bar{C}_{\mathrm{min}}}$. However NR aimed at identifying the QS, while we aim at segmenting the image, i.e. identifying plage features. For this purpose we consider a multiplicative factor, $m_p$, to the standard deviation within the disc. The contrast threshold used to identify plage is then given by the following equation: 
\begin{equation}
	\label{eq:nrfieq1}
	K=\bar{C}_{\mathrm{min}}+m_p \sigma _{\bar{C}_{\mathrm{min}}},
\end{equation}
with $m_p=8.5$.
This value was chosen so that when applied to the Rome/PSPT we replicate the SRPM \citep[Solar Radiation Physical Modelling,][]{fontenla_semiempirical_2009} plage values.
We note that the exact value of the segmentation parameter is not of great significance for the work presented here, where the aim is to compare and combine results from the various archives that have been processed consistently. 
For the segmentation, we only considered pixels within $0.98R$ for all archives to avoid uncertainties in the processing of the last few analysed pixels near the limb.

Figure \ref{fig:processedimagessamedaymask} displays examples of the derived masks of plage features for the images shown in Fig. \ref{fig:processedimagessamedayoriginal}.
These masks are used to calculate the disc fraction covered by plage regions. The same parameters were used to segment all images.
We also produced time-series of the plage areas in millionths of solar hemisphere corrected for projection effects. To do that we computed the sum of the $1/\mu$ of the plage regions, to account for projection effects, and normalised it to the area of the hemisphere.

\begin{figure*}[t]
\centering
\includegraphics[width=0.9\linewidth]{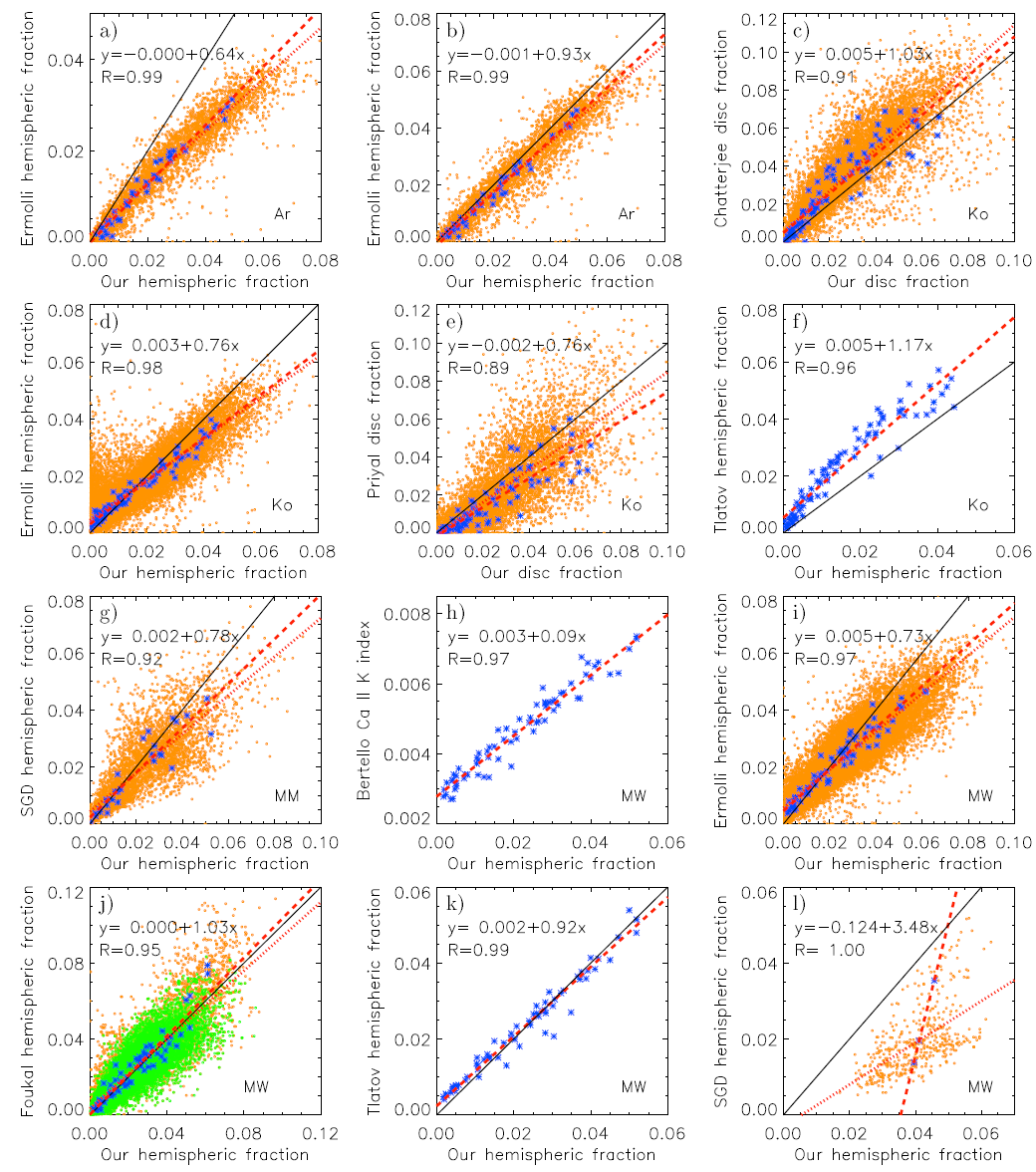}
\caption{Scatter plots between the plage area values derived by other authors ($y$-axis) and those obtained here ($x$-axis): (a)) \citet[][]{ermolli_digitized_2009} for Ar data; (b) \citet[][]{ermolli_comparison_2009} for Ar data; (c)) \citet[][]{chatterjee_butterfly_2016} for Ko data; (d)) \citet[][]{ermolli_comparison_2009} for Ko data;
(e)) \citet[][]{priyal_long-term_2017} for Ko data;
(f)) \citet[][]{tlatov_new_2009} for Ko data; (g)) SGD for MM data; (h)) \citet[][]{bertello_mount_2010} for MW data; 
(i)) \citet[][]{ermolli_comparison_2009} for MW data; (j)) \citet[][]{foukal_behavior_1996} for MW data; (k)) \citet[][]{tlatov_new_2009} for MW data; (l)) SGD for MW data. Blue asterisks (orange dots) show the annual (daily) values. In panel i) the green circles show the daily values excluding those from SC 19. The solid black lines have a slope of unity. The dashed (dotted) red lines are linear fits to the annual (daily) data. Also shown are the corresponding parameters of the linear fits to the annual values and the linear correlation coefficients of the annual values.}
	\label{fig:scatterplots}
\end{figure*}

\begin{figure*}[t]
\centering
\includegraphics[width=1\linewidth]{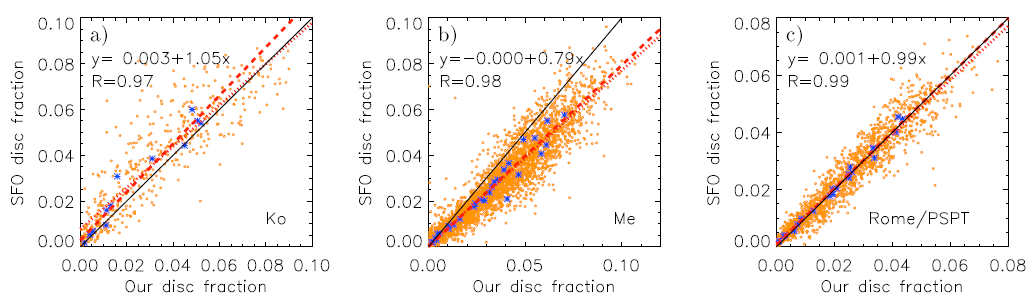}
\caption{Same as Fig. \ref{fig:scatterplots} but now comparing the plage areas derived by SFO ($y$-axis) and those obtained here ($x$-axis): (a)) for Ko data; (b)) for Me data; (c)) for Rome/PSPT data. }
\label{fig:scatterplotssfo}
\end{figure*}

\section{Analysis of the Ca~II~K series} 
\label{sec:results}

\subsection{Day-by-day comparisons between datasets}
There is, unfortunately, no single day, on which observations from all historical archives considered in this study would be available. We have identified 91 days, on which data from all historical archives other than Me, Sc, and WS are available. There is only one day of direct overlap of all archives, except Me, to either Sc or WS. Out of the 91 days on which most archives overlap we selected five days each within two days from an observations of Sc or WS to show in Fig. \ref{fig:processedimagessamedayoriginal}.
In this figure one can easily spot the differences between the archives, as well as within the same archive. The stray light conditions vary, with e.g. MM1 suffering worse conditions than Ar1.
Ko and Mi observations were taken with a similar bandwidth, which is also the broadest among the historical data analysed here. We notice that the bright features in the Ko observations are slightly smaller compared to those in the other archives. Also sunspots are seen, while Me, MW, MM, and Ar 
images display filaments which are usually not found in the data from the other archives.
The network cells and plage regions are more expanded in Ko2 compared to the other Ko observations shown in Fig. \ref{fig:processedimagessamedayoriginal}. The network cells are better seen and more expanded in Mi3 than in Ko3 even though the observations are supposed to be taken with the same bandwidth.
MM seems to be more consistent than the other archives, although it suffers more strongly from exposure problems.

In Fig. \ref{fig:processedimagessamedayflat} we notice that the contrast values of Ko images are lower than in the observations from the other archives taken with narrower bandwidth.
Therefore, we expect the plage areas derived from these data to be lower than for the other archives. 
This is not the case for observation Ko2, which might be taken with a narrower bandwidth.
We note, however, that the lower contrast can also be due to other reasons, such as stray-light, or even errors in the determination of the CC during the calibration.

The segmentation masks for the different archives shown in Fig. \ref{fig:processedimagessamedaymask} are rather similar, however some differences are noticeable. The individual regions identified as plage for the Ar, Ko Sc, and WS archives are smaller than those from MM, Me, Mi, and MW,  in agreement  with the different bandwidth of the observations and the expansion of the features with height. However, this does not apply to the Mi data, which were nominally taken with the same bandwidth as Ko observations, but suffer from clear instrumental and methodological issues. We also notice that more smaller regions are identified as plage for Mi and MW images compared to those from the other archives, as well as that the boundaries of the identified features are smoother for MM and MW than for other data sources.  
This is most likely due to different seeing conditions at the locations of the observatories or possibly due to the digitisation as well.

\subsection{Plage areas}
\label{sec:plageareas}
Figure \ref{fig:discfraction_time} shows the daily and the annual median values of plage areas in disc fractions for all the processed SHG data along with the Rome/PSPT data and the SFO series. The sunspot areas compiled by \cite{balmaceda_homogeneous_2009} are plotted in the lower panel for comparison. Considerable scatter is seen in the daily values for solar cycles (SC, hereafter) prior to SC 21. The lower scatter for SC after SC 21 is, however, also partly due to the fewer considered archives over that period. The largest dispersion (up to 0.15 in disc fraction) during the activity maximum is seen for SC 18, while plage areas during SC 19 have a smaller scatter than in SC 18. We notice the same behaviour in SC 18 and 19 also in the sunspot areas by \cite{balmaceda_homogeneous_2009}.
On the whole, the computed plage areas agree reasonably well among all datasets. There is a disagreement between the plage areas from the various datasets for cycles 17 and 20, while the plage areas from Me and MW before 1970's are higher by up to 0.03 (SC 18) than values derived from all other datasets. Me and MW include observations taken with a narrower bandwidth than most archives analysed here (only MM has a yet narrower bandwidth). This might explain the generally higher plage areas derived for those archives. The plage areas from Me match very well those from Ko for the period 1907--1914 hinting at a change in the Me data around 1914 (see Fig. \ref{fig:inconsistenciesarchives} Me1). 	MW, with a broader nominal bandwidth than Me, gives higher plage areas over SC 19 than Me. This can be due to the reported narrower slit width used in MW as compared to the other observatories \citep[see][for more details]{pevtsov_reconstructing_2016}.
Around 1980, the derived disc fractions from MW become lower than those from Ko, while the areas from Me are at a similar level to those from Ko. 
The Ko results would render SC 21 stronger than SC 19, which is in contrast to all other available data, and the sunspot areas. 
However, the results derived from the Ko series for SC 21 and 22 are less reliable than those for earlier periods. This is due to the small number of observations over that period, the lower quality of observations than for other periods, the increase of the disc eccentricities with time \citep{ermolli_comparison_2009}, and the lower accuracy of the image processing for the data after 1980 attested by the clear trend in Fig. 7b). 
Nevertheless, studies in the literature also reported stronger amplitude for some solar quantities over SC 21 than SC 19. For instance, from their analysis of Kodaikanal white light data, \cite{mandal_sunspot_2016} found that SC 21 is stronger than SC 19 when considering sunspots with area lower than 0.001 in hemispheric fraction. Our results for MM, which has a similar bandwidth  as Me, are closer to those from Ko than Me. However, the spatial resolution of MM data is lower compared to the other datasets. Furthermore, the MM images seem to suffer from poor seeing and stray-light more than other archives, which might also affect our results. 
The plage series from Mi data is not as reliable as the others due to the fact that the branch before 1957 consists of a rather small number of observations per year, commonly covering only January of each year. The lower values of Ar before SC 19 can at least partly be explained due to the change in the spectrograph on 25/05/1953  (see Appendix \ref{sec:sorting}).
The Rome/PSPT areas are systematically lower than those from Me, which is consistent with the difference in the bandwidths of their observations.

In Fig. \ref{fig:discfraction_time}, we also over-plotted the SFO values to bridge the Rome/PSPT and Ko series. These records appear to be consistent. The SFO plage areas match those from Rome/PSPT almost perfectly, while they are slightly higher than those from Ko during the maximum of SC 22.
Considering, however, that the plage areas from Ko during SC 21 and 22 are at a similar level to SC 19, this might hint to a possible underestimation of the plage areas in the Ko series before SC 21.

\subsection{Comparison of plage series from individual archives}
\label{sec:scatterplots}
Figures \ref{fig:komwcycles14-25}--\ref{fig:meothers} show the relation between the derived plage areas from the various datasets.  Following \cite{chatzistergos_new_2017} we chose one record to act as the reference and the other as the secondary one. 
We binned the disc fractional plage areas into bins of 0.005. The results do not change significantly with small variations of this value. We tested different bin sizes in the range 0.001--0.01 and found 0.01 to be too high to accurately sketch the relation very roughly, while 0.001 was too low and the bins did not include a sufficient number of data points. We produce a list of all dates when the secondary has areas within a given bin. For each such array of dates, we compute the histogram of the plage areas derived from the reference set. This results in a 2D matrix, every element of which contains the number of days for a given combination of plage areas measured by the two archives. Dividing each column by the total number of occurrences for that column results in a probability distribution function (PDF). Figures \ref{fig:komwcycles14-25}--\ref{fig:meothers} show the PDF of the plage areas of the reference archive for each bin value of the secondary one.
In these plots we also show the average values of each PDF as red circles along with the asymmetric 1$\sigma$ intervals. We applied two different types of fits to the average values (also shown in the plots), namely a linear and a power law with an offset. 
We show results for individual SC, as well as for the whole period of overlap between the compared series. 

Figure \ref{fig:komwcycles14-25} displays the relation between Ko and MW with Ko acting as the reference record. The plage areas from MW data are consistently higher than the areas from Ko data with an almost linear dependence. The results also differ slightly from cycle to cycle, although SC 18 and 21 stand out. The PDF is significantly broader in SC 18, while for SC 21 the results from the two records are in almost perfect agreement with each other. Results for SC 19 show a greater number of days with large plage areas compared to other cycles.  The distribution of MW plage areas over SC 19 shows a peak for plage areas less than 0.005 then decreases for areas 0.02--0.06 and exhibits another peak at 0.08--0.09.
We notice that the linear fit performs almost as well as the power law function function for most SC. Overall a linear relation to calibrate the results from MW to Ko series seems to be a good choice for all the cycles. 
By using different methods, \citet{tlatov_new_2009} also compared their plage areas from the Ko and MW series, by reporting a better agreement between the two series over SC 21 and a change in the relation over SC 19. However, in their results, the Ko and MW series agree also for SC 15--17.

Figure \ref{fig:mwkocycles14-25} displays the relation between Ko and MW results, now with the MW series acting as the reference record. Most aspects seen in Fig. \ref{fig:komwcycles14-25} are apparent here too, e.g. the peculiar relations for SC 18, 19, and 21, or the distribution of plage areas over SC 19.
However, the relation between the compared series now shows an obvious non-linearity, with the power law function being more appropriate to describe the relation between Ko and MW plage areas.

We obtain similar results when comparing the plage area series from the other archives, however with poorer statistics since the overlap between the compared archives is shorter.
Figure \ref{fig:komwothers} shows the same matrices as in Fig. \ref{fig:komwcycles14-25} and \ref{fig:mwkocycles14-25}, now for the Ar, MM, Me, and Mi archives and using Ko and MW as the references, due to their better temporal coverage. We notice a slight discontinuity in the disc fraction values from the Mi data compared to those from MW around 0.04, before which the areas from MW are higher than those from Mi but become lower after it. This discontinuity is also seen when compared to Ko plage areas. We investigated the reasons for this discontinuity and found that it can be attributed most likely to the instrumental changes that occurred around 1966, however we cannot exclude as probable reason the change of photographic medium on 02/03/1960.

It is worth noting that most of the relations shown 
in Fig.  \ref{fig:komwothers} are to a good extent linear. The Ar and MM data seem better fitted with the power law function for both Ko and MW as the reference, however the difference to the linear relation is minute.

In a similar manner as Fig. \ref{fig:komwcycles14-25}--\ref{fig:komwothers}, Fig. \ref{fig:meothers} compares the Me data to the other archives, now using Me as the reference.
	Note that the overlap of the Me data with the Ar, MM and Mi series is limited to roughly 700 days each, and in particular the low-activity periods are poorly represented. The overlap is somewhat better with Ko, MW and Rome/PSPT. The relationship with MW is, to a good degree, linear, while the relationship to Ko and Rome/PSPT is non-linear.
	The spread of the PDF for the comparison of Me to Rome/PSPT is lower than for the other archives. We note here that for the comparison to Rome/PSPT shown in Fig. \ref{fig:meothers} we used the Me data taken with a CCD and those stored in photographic plates. We also compared the Rome/PSPT series to the CCD and photographic plate Me data separately (shown in Fig. \ref{fig:mdpsptcycles14-25chihist}) and the results are almost the same in all cases.

Unfortunately, no historical archive, other than Me, has a statistically significant overlap with the Rome/PSPT dataset therefore we cannot perform any meaningful comparison of results from modern and the other historical observations.

\subsection{Comparison to other plage area records}

In Fig. \ref{fig:discfraction_plage} we compare annual means of our plage areas from individual archives with those available in the literature. Panels a) to e) show results derived from the Ar, Ko, MW, and MM observatories. To our knowledge, there are no published results for the Me, Mi, Sc, or WS data to compare to.
The various series are all given in hemispheric fractions except the Ko series by \cite{chatterjee_butterfly_2016}, \cite{priyal_long-term_2017}, and \cite{singh_variations_2018}, which are in disc fractions. To compare our results directly with the published series we show values of hemispheric fraction for all archives, but for Ko we also show values in disc fractions  (panel b)). 
The series by SGD, \citet[][]{foukal_behavior_1996}, \cite{ermolli_digitized_2009}, \cite{ermolli_comparison_2009}, \cite{chatterjee_butterfly_2016}, \cite{priyal_long-term_2017}, and \cite{singh_variations_2018} have daily cadence. We computed annual median values in the same way as we did for our results. The Ko series by \cite{tlatov_new_2009} is given in monthly means, while all other series are provided as annual mean values.

Ar images were previously analysed by \cite{ermolli_digitized_2009} and \cite{ermolli_comparison_2009}. Figure \ref{fig:discfraction_plage} a) shows that our series and that by \cite{ermolli_comparison_2009} agree well, while the one by \cite{ermolli_digitized_2009} is consistently lower. Our areas are at a similar level as those by \cite{ermolli_comparison_2009} after 1955, but higher before that. 
As can be seen in Fig. 7 of \cite{ermolli_comparison_2009}, their plage areas from Ar images during SC 17 are too low compared to those obtained from MW and Ko images. This supports the area values obtained here. However, the amount of images used in our study before 1950 is lower than that in \cite{ermolli_comparison_2009}, which could also contribute to the above mentioned discrepancies. 

The Ko images were analysed by \cite{kuriyan_solar_1982}, \cite{ermolli_comparison_2009}, \cite{tlatov_new_2009}, \cite{chatterjee_butterfly_2016}, \cite{priyal_long-term_2017}, and \cite{singh_variations_2018}. Our areas as well as those by \cite{tlatov_new_2009}, and by \cite{ermolli_comparison_2009} were derived from images from the earlier 8-bit digitisation of the Ko images, while the results by \cite{priyal_long-term_2017}, \cite{chatterjee_butterfly_2016}, and \cite{singh_variations_2018} are from the latest 16-bit digitisation. The results by \cite{kuriyan_solar_1982} are from the physical photographs.
Our Ko plage areas series is close to the one reported by \citet[][Fig. \ref{fig:discfraction_plage} c)]{ermolli_comparison_2009}, although it is higher for SC 21 and 22. The agreement to the series by \cite{singh_variations_2018} is also good, except for SC 14, when the areas by \cite{singh_variations_2018} are twice as high as ours. The series by \cite{chatterjee_butterfly_2016} and \cite{tlatov_new_2009} are systematically higher, while the one by \cite{priyal_long-term_2017} is almost consistently lower than ours for most SC except for SC 17, 19, and 20 (Fig. \ref{fig:discfraction_plage} b)). The values in the series by \cite{kuriyan_solar_1982} are consistently lower than ours except for SC 14. The plage areas by \cite{chatterjee_butterfly_2016} show SC 18, 19, 21, and 22 at roughly the same levels.
We can identify three potential reasons that cause the plage areas in the \cite{chatterjee_butterfly_2016} to be higher than the other series: 1) In the study by \cite{chatterjee_butterfly_2016} active regions were not excluded when calculating the standard deviation over the disc, which is then used as the threshold to segment the images; 2) bright regions were included when computing the map used to correct for the limb darkening, which reduces the contrast of plage regions \citep[cf.][]{chatzistergos_analysis_2018} and hence renders the plage regions closer to the contrast range of the QS; and 3) a threshold equal to the value of the standard deviation of the disc was used for the segmentation. We applied the same segmentation method as \cite{chatterjee_butterfly_2016} on Rome/PSPT data and found that a multiplicative factor of 1.45 to the standard deviation is needed to best match the disc fraction derived with our method.
The first two issues have unpredictable effects on the derived disc fractions, while the third one merely suggests that the method applied by \cite{chatterjee_butterfly_2016} includes fainter regions than with our method.

Figure \ref{fig:discfraction_plage} d) shows that our results from the MW series are in good agreement with those by \cite{bertello_mount_2010} and \cite{tlatov_new_2009}. We note, however, that we rescaled the series by \cite{bertello_mount_2010} to match ours by applying a linear relation. 
Our plage areas and those from \cite{bertello_mount_2010} are systematically higher than or equal to those derived by \cite{ermolli_comparison_2009} and \citet{foukal_behavior_1996}, except for SC 19. The areas by \citet{foukal_behavior_1996} are slightly lower than those by \cite{ermolli_comparison_2009}, except for SC 19 for which \citet{foukal_behavior_1996} finds considerably higher areas. The areas for MW included in the SGD series for SC 21 are lower than all other series. Overall, the results from the various studies for the MW plage areas show a better agreement than those from Ko. 

Finally, Fig. \ref{fig:discfraction_plage} e) displays that our MM plage areas agree rather well with those by SGD. Our results disagree in 1979, but our value for that year stems from merely 6 images and hence is not representative.
The series by SGD is higher than ours before 1950. This is in agreement with the comment by \cite{foukal_behavior_1996} that the plage areas by SGD for these years are inflated.

We note that \cite{priyal_long-term_2017} compared their Ko plage areas to those by \cite{foukal_behavior_1996} for MW data and found a good match between the two series, except for SC 18 and 19, for which the areas by \cite{priyal_long-term_2017} were lower by 20\%.
\cite{chatterjee_butterfly_2016} compared their Ko plage areas to MW areas by  \cite{bertello_mount_2010} and found a very good match, except during SC 19 and 22, for which the areas by \cite{chatterjee_butterfly_2016} were lower and higher, respectively.

It is worth noting that the annual values shown in Fig. \ref{fig:discfraction_plage} include all observations analysed by each study, and the data included during a given year may vary among the series.
This might also contribute to differences between the various time-series obtained here and in earlier studies.
However, the main differences come likely from the calibration, CLV compensation, and segmentation schemes applied by the various studies.
For instance, SGD, \cite{kuriyan_solar_1982}, and \citet{foukal_behavior_1996} segmented the images by manually selecting the plage regions rather than using an objective criterion. 
Artefacts in the images which were introduced by the processing or were not adequately removed, affect the results of all series. Such artefacts can appear as dark regions, reducing the recovered area of plage, or bright regions which could be misidentified as plage. 
Another critical issue is the definition of plage, which differs from study to study and is rather arbitrary.
Therefore, caution is needed even for the images for which different series give similar plage areas, since this can also be by chance.

Figure \ref{fig:scatterplots} shows scatter plots between our plage area series and those by \cite{bertello_mount_2010}, \cite{chatterjee_butterfly_2016}, \cite{ermolli_digitized_2009}, \cite{ermolli_comparison_2009}, \cite{foukal_behavior_1996}, \cite{priyal_long-term_2017}, \cite{tlatov_new_2009}, and SGD. 
For the series for which daily values are available we computed annual values for the common days only. This way we remove uncertainties due to potential differences in data selection.
The relationship between our series and those by the aforementioned authors is almost linear once we restrict the comparison to data recorded on the same days. The agreement between our results and those by \cite{ermolli_comparison_2009} for Ar, Ko, MW data and  \cite{bertello_mount_2010} and \cite{tlatov_new_2009} for MW data is good, with Person's correlation coefficients of 0.96--0.99. 

The scatter for Ko data from \cite{chatterjee_butterfly_2016} and \cite{priyal_long-term_2017}, MW data from \cite{foukal_behavior_1996}, and SGD for MM and MW data is too high to derive any meaningful conclusions. There are hints for non-linearity for the comparison to \cite{chatterjee_butterfly_2016,priyal_long-term_2017}, and \cite{tlatov_new_2009} for Ko data and \cite{foukal_behavior_1996} for MW data at high plage areas. However, for \cite{foukal_behavior_1996} this is almost entirely because of results over SC 19, when \cite{foukal_behavior_1996} reports significantly higher plage areas than all other available results (Fig. \ref{fig:discfraction_plage} d)). In Fig. \ref{fig:scatterplots} j), all daily values are shown in orange, and those excluding SC 19 are in green. The saturation is no longer evident, while the scatter is slightly reduced. We also notice that removing SC 19 slightly  increases the Pearson's correlation coefficient between the two series for daily values to 0.87 instead of 0.85. This was also reported by \cite{ermolli_comparison_2009} when comparing their results for the plage areas from MW and those by \cite{foukal_behavior_1996}.

Figure \ref{fig:scatterplotssfo} shows scatter plots between the plage areas from SFO and those derived here for Ko, Me, and Rome/PSPT. All three series show a linear relation with Person's correlation coefficients of 0.97, 0.98, and 0.99 for the Ko, Me, and Rome/PSPT series, respectively. The slopes of the linear fit are 1.05, 0.79, and 0.99 for Ko, Me, and Rome/PSPT respectively. The slopes for the fit of the Me and Rome/PSPT to the SFO series is consistent with the difference in the bandwidths of the archives. Rome/PSPT used a broader bandwidth than Me, thus resembling more the SFO series. The fit of the Ko series to SFO gives a slope greater than that of the Rome/PSPT which is contrary to expectations due to Ko having employed a bandwidth between those of Me and Rome/PSPT. This again hints for a problem with the Ko series over SC 22 (see also Sec. \ref{sec:plageareas}).

\section{Effect of image processing on derived disc fractions}
\label{discussion}

A number of plage area series have been presented in the literature. They have been derived from SHG archives using different methods. In particular, \citet{foukal_behavior_1996,foukal_extension_1998,foukal_measurement_2001,bertello_mount_2010}, and \citet{chatterjee_butterfly_2016} analysed uncalibrated SHG, while  \citet{priyal_long_2014,priyal_long-term_2017} and \citet{tlatov_new_2009} used basic calibration techniques. Also, various authors used different methods to process the data and correct for artefacts in the historical images \citep{caccin_variations_1998,worden_evolution_1998,zharkova_full-disk_2003,lefebvre_solar_2005,ermolli_comparison_2009,ermolli_digitized_2009}, while others did not do such corrections, to our knowledge \citep{ribes_search_1985,kariyappa_contribution_1996}.
Therefore, here we address the question, how the various image processing approaches affect the derived plage areas.

For this, we use part of the synthetic data created by \cite{chatzistergos_analysis_2018}, which were introduced in Section \ref{sec:syntheticdata}. In the following, we compare the plage disc fractions derived in the original Rome/PSPT observations to those from the synthetic images processed with our method and other methods presented in the literature. 
The results of this comparison for the synthetic data are summarised in Table \ref{tab:dfuncalibrated2}.

\subsection{Photometric calibration and CLV compensation}
We first compared the results from uncalibrated data to those derived from images calibrated with our method. The uncalibrated data were processed with different methods to compensate for the CLV, namely the methods by \citet[][our method]{chatzistergos_analysis_2018}, \cite{worden_evolution_1998}, \cite{priyal_long_2014}, and by using the imposed background to remove the CLV. Recall that the CC is a logarithmic function and has to be applied on the images that include the CLV in order to mimic the historical data. Using the imposed background is therefore an idealised situation where there are no errors in the image background calculation and all artefacts are removed completely. For subset 8 we also tested the calibration method by \cite{priyal_long_2014}. \cite{priyal_long_2014} applied the average CC to all images derived from the calibration wedges. Thus, we computed the CC averaged over all imposed CC, which had been randomly generated to create subset 8, and applied it on all the data. When using the calibration of \cite{priyal_long_2014} we considered two separate cases for the CLV compensation, using the imposed background and the method by \cite{priyal_long_2014}.
We did not test the calibration or CLV-compensation method by \cite{tlatov_new_2009}, because we could not replicate it with the available information. 
For all these cases we performed the same segmentation with the NR method, which was applied to obtain the results presented in Section \ref{sec:results}, while using also the same segmentation parameters employed for all the analysed data. 

The values in Table \ref{tab:dfuncalibrated2} show that the plage areas derived from the synthetic data with our processing are closest to the original ones for subsets 1 and 8, with differences being less than 0.003 and 0.025, respectively. For subset 6, the discrepancies remain quite low too, below 0.029.
It is worth noting that a linear relation was initially imposed only on the data of subsets 1 and 6, while random non-linear relations were imposed on the data of subset 8. 

The accuracy of the CLV-compensation plays an important role in the determination of the plage areas. All methods except that by \cite{worden_evolution_1998} return comparable results in the absence of strong artefacts over the disc (subset 1). However, in the presence of artefacts (subsets 6 and 8) we see increased errors in the plage areas with all of them. The methods by \cite{worden_evolution_1998} and \cite{priyal_long_2014} result in disc fractions with greater differences to the original ones than with our method. 
The method of \cite{worden_evolution_1998} performs better than the one of \cite{priyal_long_2014} in subset 8. We also note that the linear correlation coefficient for the results with the method by \cite{priyal_long_2014} is rather low ( $<0.5$) in all cases considered here except for subset 1, which is representative of the best quality historical data unaffected by instrumental and operational issues.

The CC affects the dynamic range of the contrast images in a non-linear way (remember that the CC is the relation between the logarithm of intensity and density). NR accounts for the change in dynamic range of the QS. However, errors are introduced to the plage areas due to the non-linearity affecting the plage regions. 
This can be seen in the case where we compensated the CLV with the imposed one. For this case there are no errors in the calculation of the background but the differences to the original plage areas reach 0.044 in subset 8.

\begin{figure}[t]
	\centering
	\includegraphics[width=1\linewidth]{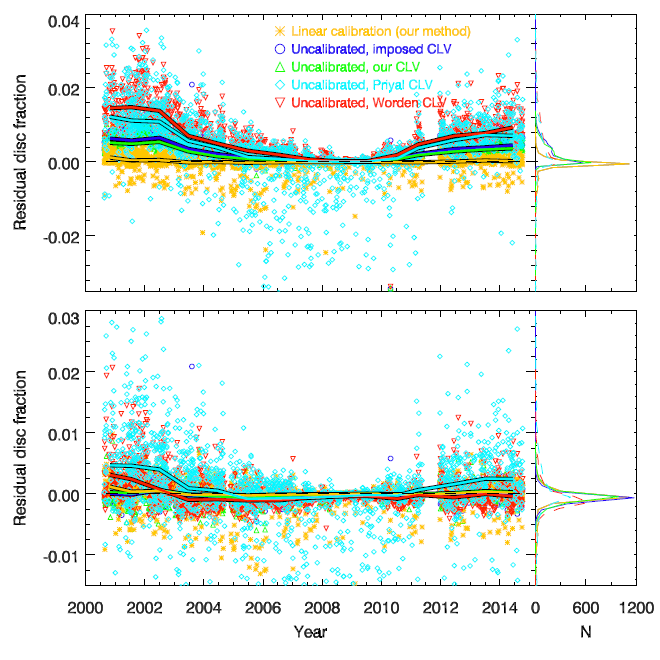}
	\caption{Left plot within each panel: Difference between the fractional disc coverage by plage obtained from the original Rome/PSPT data and from subset 8. The data from subset 8 were processed as follows: linearly calibrated with our method (yellow), uncalibrated but the imposed CLV was used to remove the CLV (blue), uncalibrated with the CLV removed with our method (green), uncalibrated with the CLV following \citet[][light blue]{priyal_long_2014}, and uncalibrated with the CLV removed following \citet[][red]{worden_evolution_1998}. The disc segmentation was done with NR using the same parameters for all cases (top panel) and by adjusting the parameters in each case (except the one for the images calibrated with our method, which is the same in both panels) to match the average disc fractions derived from all original Rome/PSPT data (bottom panel). The solid lines are annual median values of the differences in plage areas to the original Rome/PSPT data. The dashed black horizontal line is for 0 difference. Right plot within each panel: Distribution of the differences shown in the left plots.}
	\label{fig:discfractions2_residual}
\end{figure}

The top panel of Fig. \ref{fig:discfractions2_residual} shows the difference between the plage areas of subset 8 derived with the different methods and those from the original Rome/PSPT images. For all methods but ours the differences vary in-phase with the SC.

We repeated the above analysis by adjusting the segmentation parameters used in each method, except ours, so as to minimise the RMS difference in the disc fractions obtained from the various series to that from the original Rome/PSPT data. 
The results for subset 8 are shown in the bottom panel of Fig.  \ref{fig:discfractions2_residual}.
After adjusting the segmentation parameters we get comparable disc fractions with all methods, with mean differences varying between $10^{-4}$ for our calibration and $10^{-3}$ for \cite{worden_evolution_1998}. 
We notice that after adjusting the segmentation parameters the SC dependent variation is minimised from all series, however it is more evident in the results with the method of \cite{worden_evolution_1998}. 
This confirms that the method of CLV removal by \cite{worden_evolution_1998} is affected by the activity level and returns erroneous results for the active region areas.
The differences in the plage areas for subset 1 to the original Rome/PSPT data are qualitatively the same, though with lower values.

Overall, our results indicate that the photometric calibration of the images is not the most critical step for deriving the plage areas. However, plage areas from the uncalibrated data can be significantly affected by the errors in the calculation of the CLV and subsequent feature segmentation. 
Our results also indicate that it is possible to get accurate plage areas from different historical archives without the need to adjust the processing method or redefine the segmentation parameters, provided the images are accurately calibrated and have been taken with the same bandwidth.

\begin{table*}   \caption{Results from the tests performed on synthetic images.}
	\label{tab:dfuncalibrated2}   \centering   \begin{tabular}{lccccccc}
		\hline\hline
		Subset & Calibration & CLV compensation& Segmentation & \multicolumn{3}{c}{Differences} & $R$ \\
		&  & &  & Max abs. & Mean abs. & RMS &  \\
		\hline  1
		&Our method & Our method &NR & 0.003 & 0.001 & 0.001 & 0.999\\
		&Uncalibrated & Our method &NR & 0.008 & 0.002 & 0.002 & 0.997\\
		&Uncalibrated & Imposed CLV &NR & 0.010 & 0.003 & 0.004 & 0.998\\
		&Uncalibrated & \cite{priyal_long_2014} &NR & 0.010 & 0.002 & 0.003 & 0.998\\
		&Uncalibrated & \cite{worden_evolution_1998} &NR & 0.034 & 0.006 & 0.008 & 0.981\\
		&Uncalibrated & Imposed CLV &\cite{bertello_mount_2010} & 0.011 & 0.002 & 0.003 & 0.983\\
		&Uncalibrated & Imposed CLV &\cite{chatterjee_butterfly_2016} & 0.039 & 0.008 & 0.012 & 0.955\\
		\hline  6
		&Our method & Our method &NR & 0.029 & 0.003 & 0.005 & 0.942\\
		&Uncalibrated & Our method &NR & 0.026 & 0.004 & 0.006 & 0.927\\
		&Uncalibrated & Imposed CLV &NR & 0.016 & 0.004 & 0.006 & 0.979\\
		&Uncalibrated & \cite{priyal_long_2014} &NR & 0.704 & 0.014 & 0.030 & 0.186\\
		&Uncalibrated & \cite{worden_evolution_1998} &NR & 0.037 & 0.008 & 0.011 & 0.890\\
		&Uncalibrated & Imposed CLV &\cite{bertello_mount_2010} & 0.007 & 0.003 & 0.003 & 0.969\\
		&Uncalibrated & Imposed CLV &\cite{chatterjee_butterfly_2016} & 0.023 & 0.008 & 0.011 & 0.953\\
		\hline  8
		&Our method & Our method &NR & 0.025 & 0.001 & 0.002 & 0.989\\
		&Uncalibrated & Our method &NR & 0.034 & 0.002 & 0.003 & 0.993\\
		&Uncalibrated & Imposed CLV &NR & 0.044 & 0.003 & 0.004 & 0.993\\
		&Uncalibrated & \cite{priyal_long_2014} &NR & 0.755 & 0.008 & 0.026 & 0.300\\
		&Uncalibrated & \cite{worden_evolution_1998} &NR & 0.035 & 0.006 & 0.009 & 0.972\\
		&Uncalibrated & Imposed CLV &\cite{bertello_mount_2010} & 0.219 & 0.005 & 0.009 & 0.801\\
		&Uncalibrated & Imposed CLV &\cite{chatterjee_butterfly_2016} & 0.046 & 0.007 & 0.011 & 0.937\\
		&\cite{priyal_long_2014} & Imposed CLV &NR & 0.038 & 0.001 & 0.002 & 0.992\\
		&\cite{priyal_long_2014} & \cite{priyal_long_2014} &NR & 0.751 & 0.007 & 0.021 & 0.434\\
		\hline     \end{tabular}
	\tablefoot{Columns are: synthetic subset Id number, calibration, CLV compensation, and segmentation methods applied on the data, maximum absolute differences, mean absolute differences, and RMS differences, as well as the Pearson coefficient between the disc fractions calculated in each case and those derived from the original Rome/PSPT images.}
\end{table*}

\begin{figure}[t]
	\centering
	\includegraphics[width=1\linewidth]{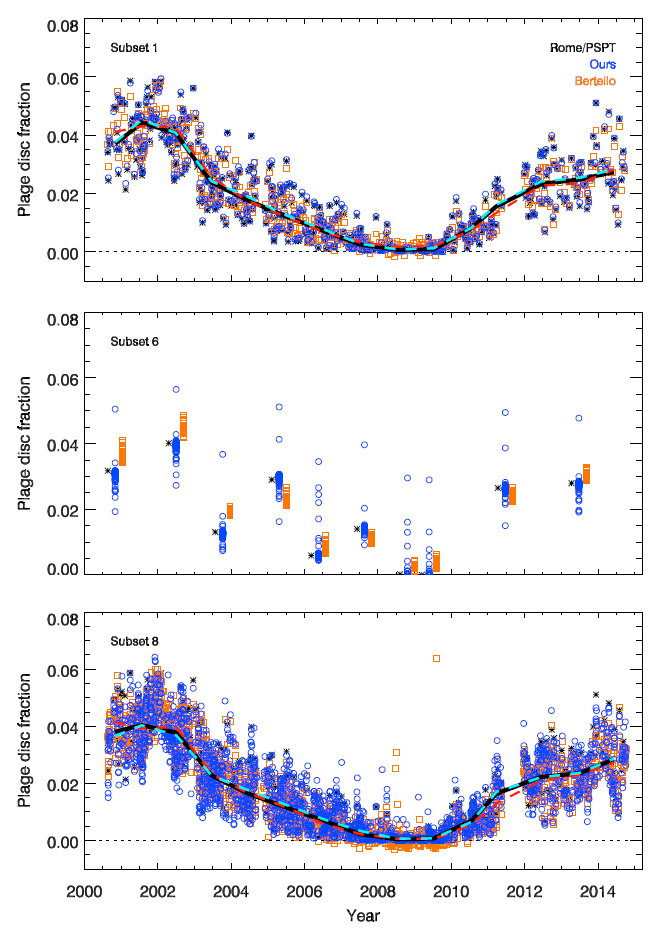}
	\caption{Comparison between plage area disc fractions derived with our method (blue circles) and that by \citet[][orange squares]{bertello_mount_2010} for subset 1 (top panel), subset 6 (middle panel), and subset 8 (bottom panel). The disc fraction from the original Rome/PSPT data used to create the synthetic images are shown in black asterisks. Notice that the blue circles lie almost perfectly over the black asterisks for subset 1. To improve the visibility of the results for subset 6, the plage areas derived with our method and those by \citet[][]{bertello_mount_2010} have been shifted in the $x$-axis by 0.2 and 0.4 of a year, respectively. Annual median values are shown for the original Rome/PSPT data (solid black line) and for the synthetic data processed with our method (dashed light blue line) and \citet[][dashed red line]{bertello_mount_2010}. The dotted black horizontal line is for plage area disc fraction of 0.
		Note, subset 6 is available only for 10 days during the considered period. The negative values of plage areas obtained with the method by \citet[][]{bertello_mount_2010} are due to the linear scaling applied to match the results to the original Rome/PSPT plage areas, see Sec. \ref{sec:bertechat} for more information.}
	\label{fig:bertelloplagebplage}
\end{figure}

\subsection{Image segmentation} 
\label{sec:bertechat}
Here we test the effect of segmentation approaches on the derived plage areas.
We do not intend to do a comprehensive test of different segmentation approaches, but simply compare the method employed here and the two methods most commonly used with historical Ca~II~K data. These are the method developed by \citet{bertello_mount_2010}, and the method using a multiplicative factor to the standard deviation of the values within the disc \citep{foukal_extension_1998,chatterjee_butterfly_2016}.

\citet{bertello_mount_2010} presented a method to derive a plage area index from uncalibrated Ca~II~K observations.
They produced density contrast images by dividing the images with a 2D map resulting from applying a running window median filter to the image. Then a Gaussian function was fit to the histogram of the whole image. A second histogram was calculated, this time keeping only the regions within $x^{+7\sigma}_{-2\sigma} $, where $x$ is the centre of the Gaussian. The histogram was split into 30 bins and divided by the total number of pixels. Another four-parameter Gaussian was fitted to these normalised bins, and the Ca~II~K index was defined as the additive parameter from the fit.
We studied the accuracy of this index with the synthetic data and compared results with those derived with our processing of the same data.
Since \cite{bertello_mount_2010} do not provide adequate information on how the MW images were compensated for the QS CLV (e.g. size of window width of median filter), we used the imposed CLV to remove it so that in this process we had no errors due to miscalculation of the image background. Therefore, the errors we find can only be considered as a lower limit of the error introduced by the image processing of \citet{bertello_mount_2010}. In this test, we noticed that the choice of the authors to use a fixed bin size of 0.01 in contrast for the calculation of the first histogram resulted in merely 4 points for 12 days during quiet periods. This issue was not mentioned by \citet{bertello_mount_2010}, therefore we decided to simply ignore these days for our comparisons here.
The values derived with the method by \citet{bertello_mount_2010} were then linearly scaled using parameters of the linear fit to the plage areas from the original Rome/PSPT data.
In this way we achieve on average the best match with the original Rome/PSPT values. However, we obtain negative values of disc fractions for some of the data, which is unrealistic. Requiring all values to be positive reduces significantly the agreement between the two series, however.  

Figure \ref{fig:bertelloplagebplage} shows plage regions derived from all synthetic data of subsets 1, 6, and 8 with our method and that by \citet{bertello_mount_2010}. 
For subset 1 and 8, the differences between the Ca~II~K index by \citet{bertello_mount_2010} and the original data are greater than those for our processing, but they are lower for subset 6. 
This shows that the method by \citet{bertello_mount_2010} can return consistent results in the case when the image background is calculated with high accuracy and the CC is constant with time.
For subset 8, where the CC varies among images, the errors with the method of \citet{bertello_mount_2010} increase up to 0.22 for a few images during activity minimum. Overall, the results with \citet{bertello_mount_2010} are comparable to those with our method, but calibrating the observations gives yet more accurate results. 
However, it should be noted that part of the differences might be due to different definitions of the contrast threshold of the plage regions. Considering that the Ca~II~K index by \citet{bertello_mount_2010} does not reach a plateau during activity minimum, it might partly include the network component.
We repeated these calculations by forcing the fit between the \citet{bertello_mount_2010} Ca~II~K index and the original Rome/PSPT plage areas to have only positive values. The agreement between the Ca~II~K index values and the plage areas worsened, with RMS differences doubled for all subsets.

We also compared the plage areas derived with a segmentation method similar to that of \cite{chatterjee_butterfly_2016}, when applied to uncalibrated images and flattened with the imposed background. The segmentation is performed by using a threshold, $K$, in the form:
\begin{equation}
	\label{eq:mlffieq}
	K=\textrm{  median}(C)+m \sigma,
\end{equation}
where $m$ is a multiplicative factor and $\sigma$ is the standard deviation of contrast values within the disc. 
\cite{chatterjee_butterfly_2016} also applied a morhological closing operator to the segmentation mask, which we did not apply. 
The errors with this method are slightly lower than those with our processing for subset 6, but are much higher for subsets 1 and 8 (Table \ref{tab:dfuncalibrated2}). This shows that this method is less versatile in its response to varying CC than ours.

The results presented here are in agreement with those of \cite{chatzistergos_analysis_2017} concerning a similar and less extensive test for the effects of photometric calibration on the plage areas derived from application of a constant contrast threshold on the images.

\section{Composite series}
\label{sec:composite}
In this section we use the plage areas derived from the different archives, which were presented in Section \ref{sec:plageareas}, to combine them all into a single record. Since Ko and MW are the longest and most homogeneous series, we produced two separate cross-calibrated series by using Ko and MW sets as the reference. However, as shown in the previous sections, neither of the considered series is free of problems. Even though Me is the longest series considered in our study, it could not be used as a reference due to the low number of data over extended periods and different set-ups employed for the digitisations.
Moreover, due to the short overlap with the historical archives, we cannot use the plage series from the modern Rome/PSPT observations as the reference.

\begin{figure*}[t]
	\centering
	\includegraphics[width=1.0\linewidth]{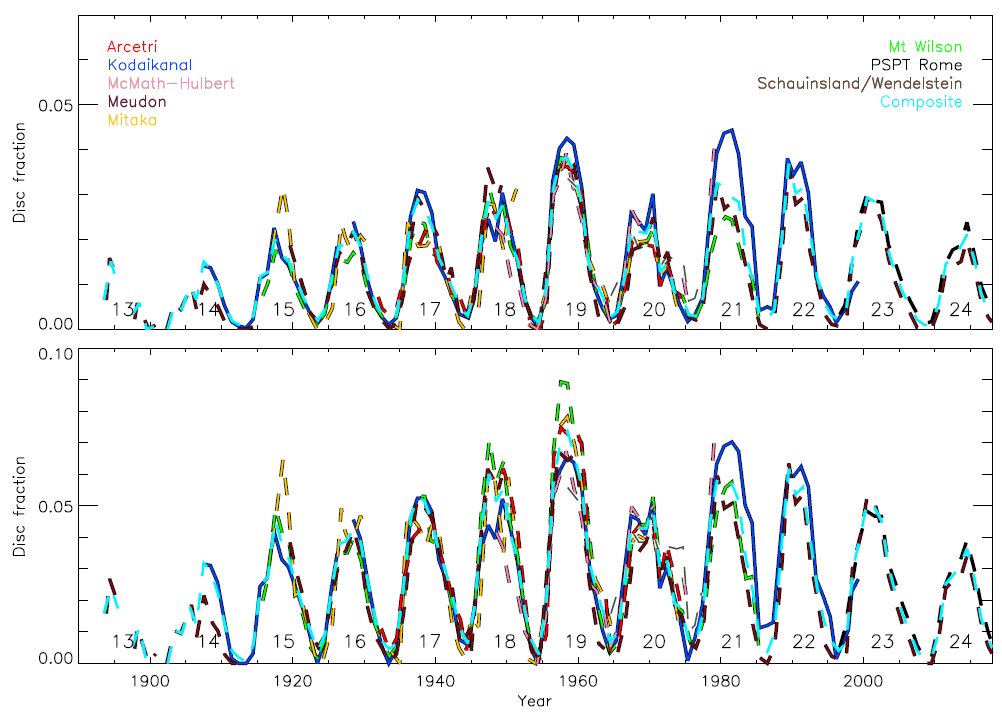}
	\caption{Fractional disc coverage by plage as a function of time, derived by applying the NR method with the same parameters on the images from the Ar (red), Ko (blue), MM (pink), Me (brown), Mi (orange), MW (green), and Sc/WS (dark green) archives after calibration to the plage areas from the Ko (top) and MW (bottom panel) series. Individual small dots represent daily values, while the thick lines indicate annual median values. The dashed light blue line is the composite of all calibrated plage series, while the solid black curve is the unscaled plage area series from Rome/PSPT. The numbers under the curves denote the conventional SC numbering.}
	\label{fig:discfraction_time_calibratedKoMW}
\end{figure*}

\begin{figure*}[t]
	\centering
	\includegraphics[width=1.0\linewidth]{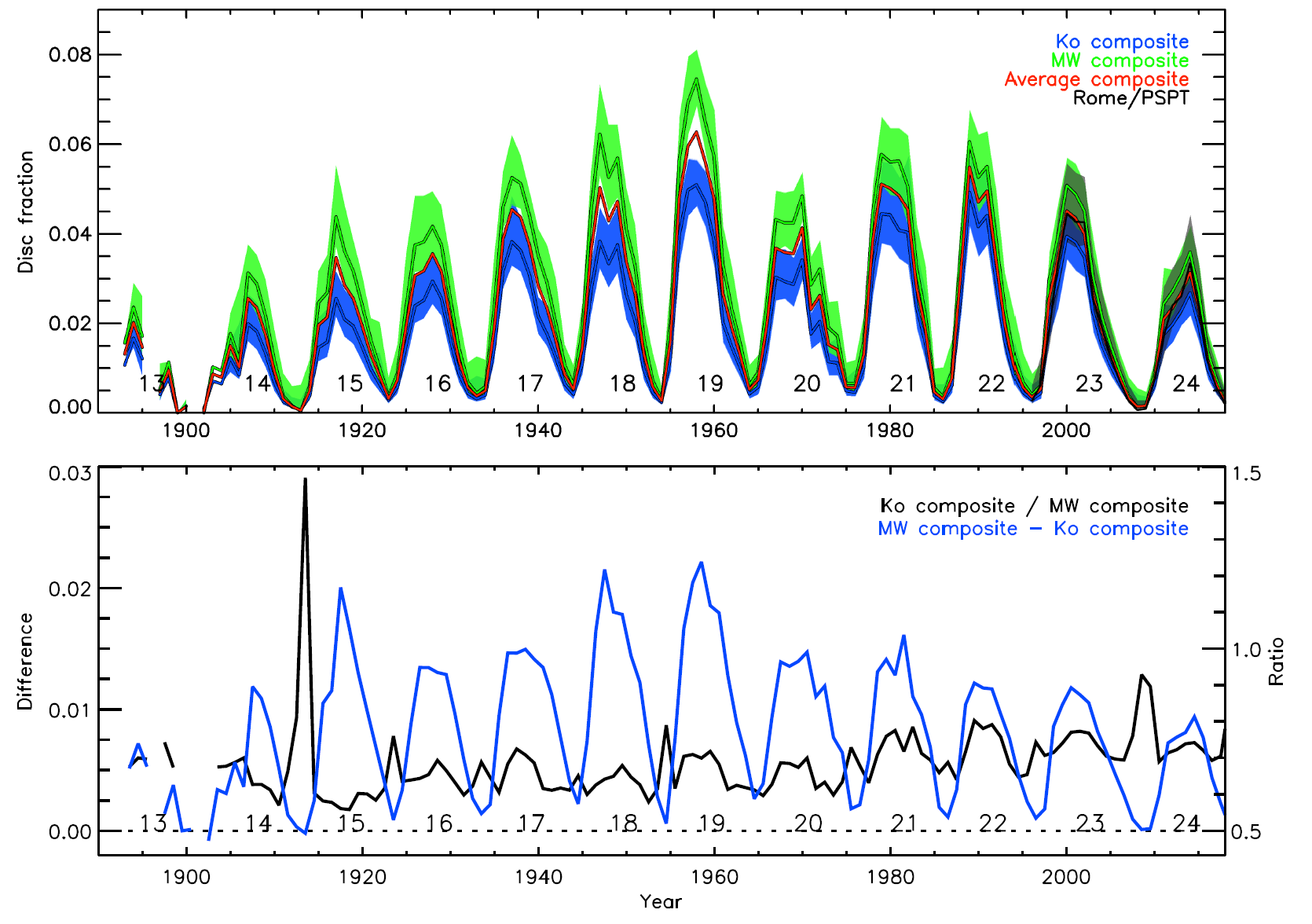}
	\caption{Composites of fractional disc coverage by plage as a function of time (upper panel), by using the data from Ko (blue) and MW (green) as the reference. Also shown is the average of the two composites (red), and the unscaled plage areas from Rome/PSPT (black). The shaded surfaces give the asymmetric $1\sigma$ interval, while the solid lines are annual median values. The lower panel shows the ratio (black) and difference (blue) between the annual values of the composites based on the Ko and MW series.  The dotted horizontal line denotes the ratio of 0.5 and difference of 0 in disc fractions. The numbers under the curves in both panels denote the conventional SC numbering.}
	\label{fig:discfraction_time_composite}
\end{figure*}

We used the relations between the archives presented in Section \ref{sec:scatterplots} for the entire period of overlap between the various data sources. 
The data from Ar were split into two separate series, those before and after 25/05/1953 in order to take into account the instrumental change that occurred on that date (see Sec. \ref{sec:sorting}). 
The data from Mi were also split into two segments, those before and after 01/02/1966 to account for the instrumental changes. We did not consider the discontinuity in the Mi series at the end of 1924 to account for the relocation of the observatory, because of the very low number of images before 1924. 

The MW data were also split before and after 21/08/1923 to account for the change to that instrument.
For Ar, Mi, and MW we derived the relations for each segment separately. 
When calibrating the different archives to Ko series, we considered the linear relations, which in most cases provide a good approximation of the relationships between the series (see Fig. \ref{fig:komwcycles14-25} and \ref{fig:komwothers}), and which is more stable to extrapolate to larger  plage coverages for which only few data points are available. Four exceptions were made, for Ar, MM, Me, and Mi data before 01/02/1966, for which we used the power law function which was found to fit the data best. The power law function was also employed to calibrate the results from all archives to the MW series, as the corresponding relationships are clearly non-linear (Fig. \ref{fig:mwkocycles14-25} and \ref{fig:komwothers}). 

We produced a composite record for each reference archive by appending the results from all calibrated series to those from the reference dataset. 
The unscaled MW data before 21/08/1923 were included in the MW composite, however they were not used to derive the relations between the MW plage areas and those from the other observatories.
Daily mean values of the composites were calculated and from those monthly and then annual mean values.
The composites (only the annual values) are shown in Fig. \ref{fig:discfraction_time_calibratedKoMW}, along with the scaled individual series entering them. The unscaled annual median plage areas from Rome/PSPT are also shown. The Rome/PSPT values were also included in the composites by considering Me data. In particular, we used the derived relation to calibrate Rome/PSPT areas to those from Me and then the obtained relations to calibrate Me to either the Ko or MW series.

The agreement between the series of individual data sources has improved for both reference archives compared to the uncalibrated data, with the exception of the results obtained for SC 21. The RMS difference of the daily plage areas derived from the various archives to those from Ko and MW is reported in Table \ref{tab:rmsdifferenceplageareas}. 
We notice that the Ko based composite differs to the individual Ko plage series in that SC 19 and 21 have reduced amplitude by up to 0.008 and 0.024 in the annual values, respectively. Similarly, the main difference between the MW based composite to the individual MW plage series is the reduced amplitude in SC 18 and 19 by up to 0.009 and 0.016 in the annual values, respectively.

Figure \ref{fig:discfraction_time_composite} shows annual median values of the two composites (top panel),
as well as the ratio between the two composite records (bottom panel).
These two composites include 35,094 out of 45,655 days, which is the period covered by Me, thus achieving 77\%~coverage.
The two composites show a very similar evolution of plage areas, with an almost constant increase over SC 15--19, a drop over SC 20 to similar levels as SC 16 and an increase again over SC 21 and 22, followed by a decrease over SC 23 and 24.
The composite based on Ko plage areas has a lower overall level  than the one 
derived using MW as reference. The plage areas in the MW based composite become lower than those from the Ko based composite only for the solar activity minima around 1902 and 1913 by 0.0009 and 0.0002 in disc fraction, respectively. The difference over the whole period of time remains less than 0.059 and 0.024 for the daily and annual values, respectively. SC 19 is the strongest in both composites, although for the Ko composite it is only slightly stronger than SC 22.  
SC 15 is slightly lower (by 0.004) than SC 16 for the Ko composite, while for the MW composite SC 15 is higher (by 0.002) than SC 16. 
SC 17 and 18 have the same amplitude for the Ko composite, while for the MW composite SC 18 is higher (by 0.01) than SC 17. 
The ratio of the Ko to MW composites at SC maxima remains roughly constant at $\sim0.66$ over SC 15--20, but increases to $\sim0.75$ for SC 21--24.

It is worth noting that the unscaled Rome/PSPT areas match exceptionally well the average composite. This is consistent with the scaling of the various series we have presented in Sec. \ref{sec:scatterplots}. Scaling Rome/PSPT to Me requires a multiplicative factor of 1.20 (Fig. \ref{fig:meothers}), while Me to Ko and MW a factor of 0.75 and 0.92 (Fig. \ref{fig:komwothers}), respectively. This gives the coefficient 0.9 and 1.1 to calibrate Rome/PSPT to Ko and MW, respectively, with an average value of unity.

The differences in the absolute levels of the two obtained composites can be explained to some extent by the diverse bandwidths of the two reference archives. Ko and Mi are the historical datasets with the broadest wavelength bands and show lower disc fractions for most of the period. MW with its narrow bandwidth samples higher regions in the solar atmosphere where the plage and network regions cover a larger area, while the sunspots are diminished. Hence, sunspots, which are dark in Ko data, will appear, to some extent, as plage in MW data. 

We notice that the two composites lie apart by roughly 1$\sigma$ of each series over SC 15--20.
	The above differences in the obtained composites illustrate the uncertainty in the results due to the selection of reference datasets. In Fig. \ref{fig:discfraction_time_composite} we also show the average series of the Ko and MW based composites. This is our proposed preliminary plage areas composite series\footnote{Available at \url{http://www2.mps.mpg.de/projects/sun-climate/data.html}}, which we intend to keep updating with new data and with improvements in the methodology used to create it.

For example, we note that cross-calibrating the series with average parameters for their entire period carries errors due to inconsistencies within the considered archives themselves.
These inconsistencies need to be identified and, if these are isolated images, to be separated from the rest. In the case of documented instrumental changes (like the ones in Ar, Mi, or MW) we suggest to split the affected datasets and treat each part as a separate archive, as done in this study.
Furthermore, we processed and segmented all archives in exactly the same way. This allows to detect inconsistencies between the different archives, but it might not be the optimal method to produce a composite from the available datasets due to the differences in their bandwidths.

\begin{table}
	\caption{RMS differences between the daily plage areas derived from all analysed datasets to those from the Ko and MW series before and after their calibration. }
	\label{tab:rmsdifferenceplageareas}
	\centering
	\begin{tabular}{l*{4}{c}}
		\hline\hline
		\small
		&\multicolumn{2}{c}{Uncalibrated} &		\multicolumn{2}{c}{Calibrated} \\
		&Ko&  		MW&    		  Ko& 			MW\\
		\hline
		Ar (before 29/05/1953) & 0.012 & 0.022 &0.012 & 0.014  \\
		Ar (after 29/05/1953)		   &0.015  & 0.014 &0.011 &0.012\\
		Ko &0.000  & 0.018 &0.000 & 0.013\\
		MM &0.013  & 0.019 &0.011 & 0.016\\
		Me &0.016  & 0.010 &0.011 & 0.010\\
		Mi (before 01/02/1966) &0.017  & 0.021 &0.013 & 0.017 \\
		Mi (after 01/02/1966) &0.025  &0.014  &0.011 &0.010\\
		MW (before 21/08/1923) & 0.020 &0.000  &0.007 & 0.000\\
		MW (after 21/08/1923) & 0.018 &0.000  &0.011 & 0.000\\
		Sc/WS & 0.015 & 0.019 &0.014 & 0.015\\
		\hline
	\end{tabular}
	\tablefoot{The first two columns with RMS differences are for the uncalibrated series, while the last two are for the calibrated series as described in Sect. \ref{sec:composite}.}
\end{table}

\section{Conclusions}
\label{sec:conclusions}
We have processed 8 historical Ca~II~K SHG archives and a modern one from the Rome/PSPT to derive the disc coverage by plage regions.
We processed the data with the method we developed in \cite{chatzistergos_analysis_2018} and calculated  plage areas over the whole period the data are available, from 1893 to 2018. 

We showed that accurate processing of different historical SHG time-series is possible, which in turn allows consistent results in terms of plage areas. We found that the photometric calibration is not the most critical step if the aim is only to derive plage areas. The most important step is an accurate estimate of the background of the image (i.e. the quiet Sun CLV) and accounting for image artefacts.  
However, accurate calibration improves the accuracy of the plage area estimates. We evaluated other methods of processing uncalibrated data, and showed 
that the plage areas obtained after applying our method are closer to those one would get for ideal data (represented here by modern Rome/PSPT observations) than other methods in the literature allow.

We also showed that processing the data with our method does not require an arbitrary adjustment of the segmentation parameters depending on the archive, as needed for all other methods.

Analysis of a large set of historical archives allowed us to detect inconsistencies in the images owing to instrumental/observing issues, respectively.
We identified instrumental changes that affect the plage areas obtained from the archives of Ar, Mi, and MW. In addition, the quality of Ko data was found to change significantly with time. However, more work is needed to identify all possible instrumental and methodological changes affecting the series and in particular to correct for these.
	
We compared our derived plage areas to those published in the literature we showed that there are inconsistencies in the various series owing to unaccounted instrumental/observing or processing/methodological issues.

We used the two longest and most stable records from Ko and MW observatories as the reference to produce two composites of plage areas from all analysed archives. Based on the available Me data we could combine results from both historical and modern observations. We found a linear relation between the plage series to be a reasonable approximation when Ko plage area series is considered as the reference. However, a power law relation seems more appropriate when the MW plage area series is the reference.
The two composites differ in the absolute level of the plage areas, with differences that depend on the SC and SC phase. The two composites show different long-term trends. This highlights the importance to understand all instrumental/operational changes affecting the series before drawing conclusions on long-term solar trends from these data. Nevertheless, by averaging the two obtained composites, we produced our preliminary plage area record covering the period 1893--2018. To our knowledge, this is the first composite of plage areas derived by the integration of results from several archives.

Finally, it is worth noting that despite the large number of archives and images analysed in our study, there are 10561 days with no data in the produced composite series. Besides, clear inhomogeneities affect most of the analysed observations. Inclusion of more archives, e.g. those from Catania, Coimbra, Kyoto, Rome, Sacramento Peak, and of data from new digitisations, e.g. Kodaikanal,  can help fill these gaps. Furthermore, identification of homogeneous data is important for the creation of an accurate series of long-term variation of plage areas.
Future work will aim at this in order to improve the produced composite series.

\begin{acknowledgements}
	We thank Takashi Sakurai for providing us with the Mitaka data. We kindly thank Hubertus W\"ohl for having scanned and provided to us the digitised images from Schauinsland and Wendelstein observatories. We thank Subhamoy Chatterjee, Angie Cookson, Muthu Priyal, and Andrey Tlatov for providing time-series of plage areas.
	We thank Isabelle Buale, Jean-Marie Malherbe, and Brigitte Schmieder for providing valuable information and help about the Meudon data.
	The authors thank the Arcetri, Kiepenheuer Institute for Solar Physics, Kodaikanal, McMath-Hulbert, Meudon, Mitaka, Mount Wilson, and the Rome Solar Groups. We also thank the anonymous referee for the constructive comments that improved this paper.
	T.C. and I.E. thank the International Space Science Institute (Bern, Switzerland) for supporting the International Teams 417 "Recalibration of the Sunspot Number Series" and 420 "Reconstructing solar and heliospheric magnetic field evolution over the past century", respectively. 
	T.C. acknowledges postgraduate fellowship of the International Max Planck Research School on Physical Processes in the Solar System and Beyond.
	This work was supported by grants PRIN-INAF-2014 and PRIN/MIUR 2012P2HRCR "Il Sole attivo", COST Action ES1005 "TOSCA", FP7 SOLID, and by the BK21 plus program through the National Research Foundation (NRF) funded by the Ministry of Education of Korea. This research has made use of NASA's Astrophysics Data System.
\end{acknowledgements}

\bibliographystyle{aa}
\bibliography{_biblio1}   
\clearpage
\appendix

\section{Data presentation and sorting}
\label{sec:sorting}
\begin{figure}[t]
\centering
	\includegraphics[width=1\linewidth]{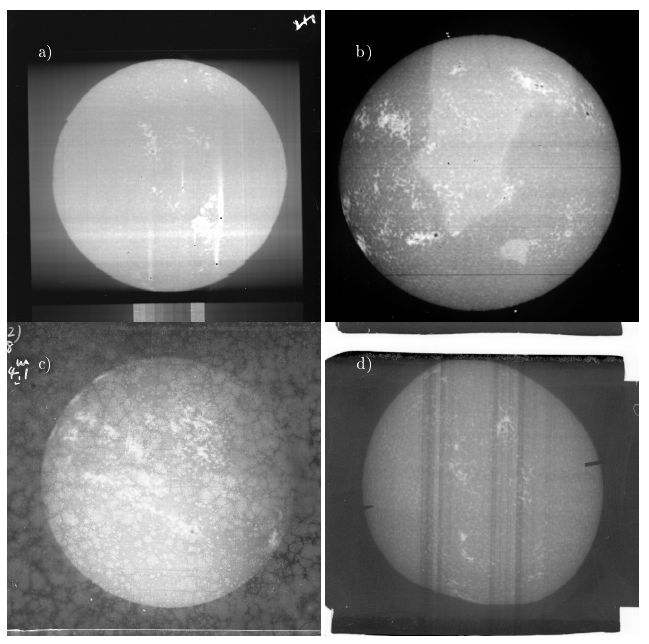}
\caption{Examples of historical observations that were excluded from this study. Shown are the raw density images. The plotted examples were 
taken at the: (a)) Arcetri (21/03/1941), (b)) Kodaikanal (18/03/1990), (c)) Mitaka (11/04/1948), and (d)) Mt Wilson (17/01/1973) observatories. The images are not compensated for solar ephemeris.}
	\label{fig:01_01_shg_examplesexcluded2}
\end{figure}

\begin{figure}
	\centering
	\includegraphics[width=1\linewidth]{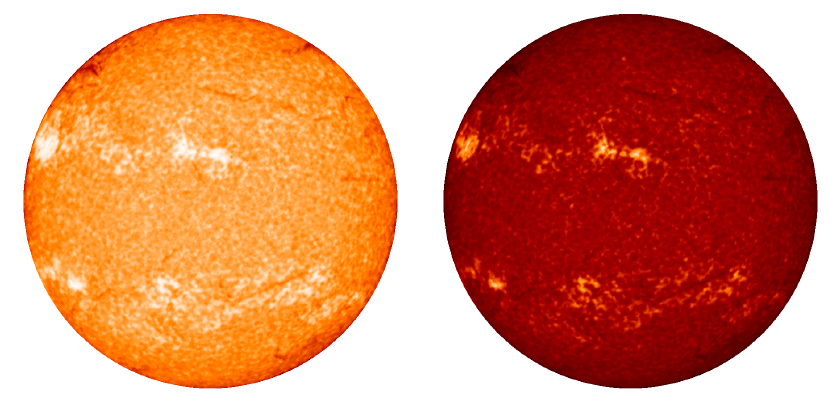}
\caption{Me observation taken on 30/10/1998. Left: image converted to density values automatically by the scanner; right: image after our correction.}
	\label{fig:19981030densitytheirs}
\end{figure}

Many historical images suffer from various problems that do not allow any meaningful analysis. Images with high noise levels,  extremely low contrast regions (hinting at exposure problems), missing parts of the disc, severely distorted disc, or artefacts that look like plage were excluded from this work. This resulted in 16904 images being rejected from all archives combined. Examples of such images are shown in Fig. \ref{fig:01_01_shg_examplesexcluded2} and in \cite{chatzistergos_analysis_2017}. It should be noted that for this study we did not perform any elaborate data selection to homogenise the datasets, we rather ignored pathological cases or images for which the assumptions made during our processing break down, due to e.g. severe exposure problems. 
For instance, a large number of images from the Mi archive exhibit problems which impede automatic and accurate processing, see e.g. Fig. \ref{fig:01_01_shg_examplesexcluded2} c), while many observations over the period 1917--1949 also exhibit significant overexposure problems. Therefore, we could only use less than 50\% of the available data from Mi before 1950. 
Furthermore, some active regions in the Ar, Ko, Me, Mi, and WS data are saturated. This seems to have resulted from the digitization and is seen as a single cut-off transparency value. 
The information we can derive from such data is limited. However, if the saturation was not extended to the network regions, our code for the image processing tagged the saturated regions as plage regions and used such images to gain information on fainter features. 
This is not applicable to the Me data for which the image saturation is not characterised by a single cut-off value. The saturation in the Me data seems to have been introduced, at least partly, by the conversion of transparency values to density values during the digitisation, as the conversion was done automatically by the scanner with a linear relation. We applied a linear relation to revert the image to its negative and then convert the values to densities based on the photographic theory $d=\log{(1/T)}$, where $T$ are the transparency values \citep[see][]{dainty_image_1974,chatzistergos_analysis_2018}. Figure \ref{fig:19981030densitytheirs} includes one observation from Me before and after our correction for the saturation. However, we point out that the use of these data is limited to the scope of this study focusing on deriving plage areas and can't be used to study the contrast of the plage regions.

Some images from all archives show a distorted solar disc. These distortions can be due to irregular motion of the slits of the spectroheliograph over the solar disc during the observation or due to a similar issue during the digitisation with a linear array, as was used for the 8-bit digitization of the Ko data. These problems would cause strips of the images to be stretched or compressed. Such distortions introduce inherently large errors in the derived plage areas. Therefore, severe cases of disc distortions were excluded from our analysis.

\begin{figure*}[t]
	\centering
	\includegraphics[width=1\linewidth]{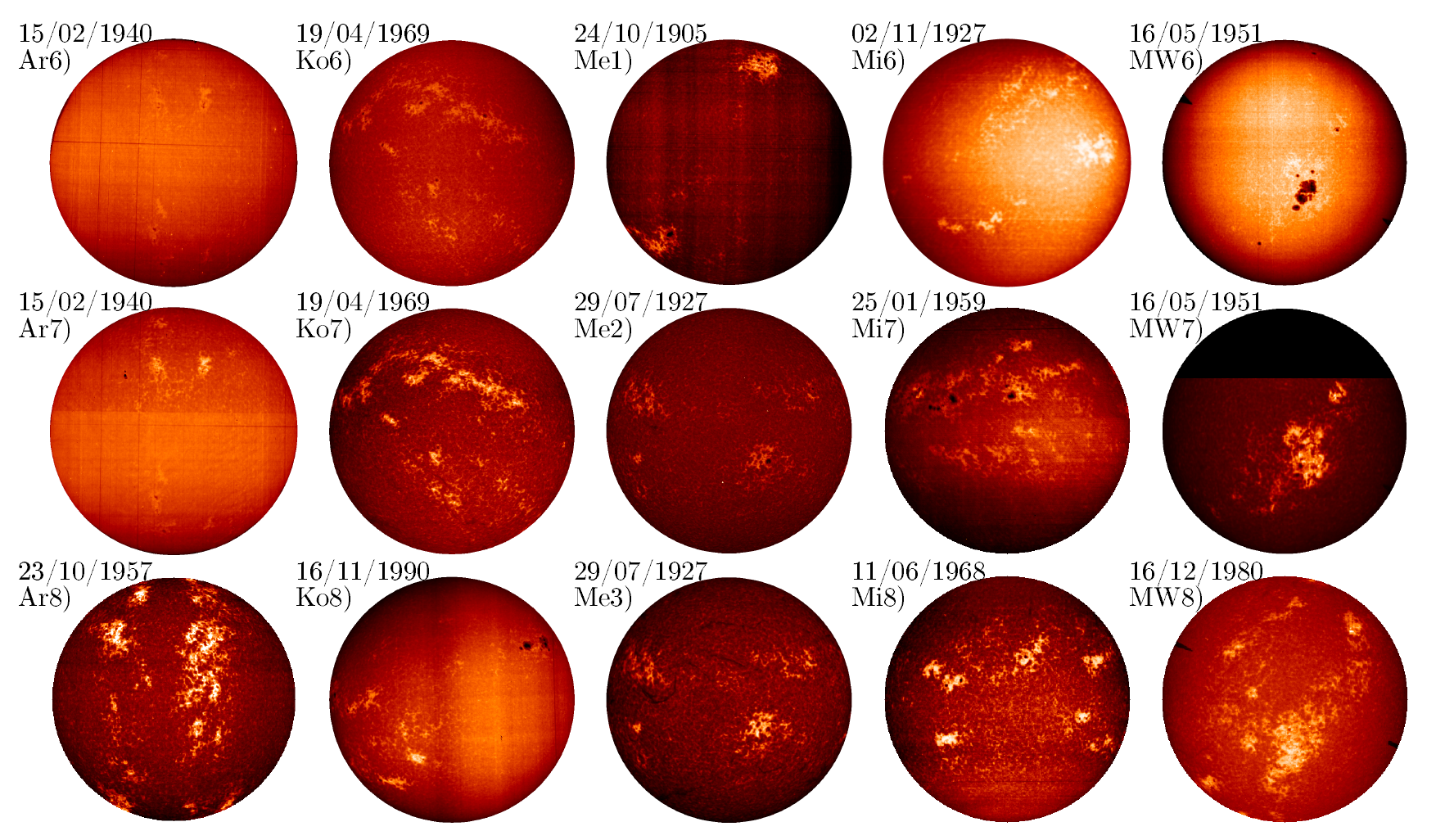}
\caption{Examples of the diverse images included in the considered historical Ca~II~K archives to show the different atmospheric heights sampled by the various observations. Shown are raw density images from the archives of Ar, Ko, Me, Mi, and MW. The images are saturated to the range of values within the solar disc. The images are not compensated for ephemeris. Unlike in Fig. \ref{fig:processedimagessamedayoriginal}, \ref{fig:processedimagessamedayflat}, and \ref{fig:processedimagessamedaymask} the images from the different observatories do not correspond to the same date. }
	\label{fig:inconsistenciesarchives}
\end{figure*}

While analysing the data, we found a lot of duplicate files in the MW series by applying a code that checked for the pixel-by-pixel similarity of the images. Most of these pairs of files listed the same observation days, while more than 80 pairs of files contained the same image, but listed different dates. The date of these recordings was identified by comparing with the other analysed datasets. The duplicate images with the wrong date were excluded from our analysis. Other datasets sometimes have incorrect dates too,  due to mistakes in naming files during the digitisation, and archiving and retrieving procedures. These kinds of errors in the data are unsurprising considering that the digitisation process (which was not limited to Ca~II~K images) had to deal with $\sim10^4-10^5$ files, usually lasted several years, and was often performed for multiple observations simultaneously (for instance two, three, or even four plates in a single image). This highlights the need to analyse multiple SHG series to catch and mitigate these errors.

It should also be noted that the bandwidths reported here for each observatory are only the nominal ones, while the actual bandwidths may vary significantly over time, as well as even within one observation \citep[see][]{chatzistergos_analysis_2018}. 
Instrumental changes can be the reason for some of these variations.
For example, the Ar logbook states that in 25/05/1953 the second slit of the spectrograph was moved closer to the photographic plate, thus making the effective bandwidth of the observation narrower.
Figure \ref{fig:inconsistenciesarchives} shows additional raw images from the Ar, Ko, Me, Mi, and MW archives to illustrate the diversity of images included in these data sources. Notice that these images do not correspond to the same dates. We found images with significant photospheric contributions in all archives, see e.g. panels Ar6, Ko6, Mi7, MW6 in Fig. \ref{fig:inconsistenciesarchives}. For instance, observations shown in Fig. \ref{fig:inconsistenciesarchives} Ar6 and Ar7, MW6 and MW7, as well as, Ko6 and Ko7 were taken on the same day, but clearly sample different atmospheric heights. Thus, while sunspots are barely visible in most MW images of Fig. \ref{fig:processedimagessamedayoriginal} and \ref{fig:inconsistenciesarchives}, they are very clearly seen in Fig. \ref{fig:inconsistenciesarchives} MW6 and in many other images of this series. Similarly Me2 and Me3 are taken on the same day, but filaments are seen only in Me3, while Me2 shows sunspots, suggesting that Me2 was not centred at the core of the line. There are also images for which a change in the bandwidth or the central wavelength occurred during the observation, e.g. Fig. \ref{fig:inconsistenciesarchives} Ar7 and Ko8.
Furthermore, we found that observations taken in different lines, such as H$\alpha$ or white light have been mixed with the Ca~II~K ones during the digitisation of Ko, MM, Me, and MW archives. Ca~II~K observations were also found within the H$\alpha$ archive from the MM observatory.
Hence, the digital series from SHG historical archives include a collection of diverse observations with different quality. Images that were clearly taken in a different spectral range were removed from our analysis. Since several of the analysed archives are available on public repositories, it is worth noting all of the above issues affecting these data to highlight the need for a very careful study in order to derive accurate results from them.

\section{Pre-processing}
\label{sec:preprocessingstep}
In the first step we removed small artefacts and cleared outliers in the observations, due to dust or holes in the photographic plate. To this purpose, we applied a low-pass filter on the image with the width of 5 pixels, produced a contrast image, $C_m$, by dividing the original image with the one resulting from the low-pass filter, and replaced the pixels of the original image at the locations where $C_m>1.2$ or $C_m<0.5$ with the corresponding values from the smoothed image. 
These thresholds were chosen very conservatively, in order to ascertain that even the brightest plage or the dark parts of the sunspots are not affected by this procedure.

Estimates of the centre of the solar disc and the radius are listed in the FITS file headers for the Ar \citep{ermolli_digitized_2009}, Ko \citep{makarov_22-years_2004}, Me \citep{zharkova_full-disk_2003}, and MW \citep{lefebvre_solar_2005} archives, but not for the rest. 
Deriving these parameters is a complicated task, because historical SHG observations very rarely show a regular solar disc \citep[see][or Fig. \ref{fig:01_01_shg_examplesexcluded2}d)]{ermolli_comparison_2009}, due to e.g. distortions or misalignments of subsequent slit observations. 
Besides that, artefacts affecting the image near the limb and outside the disc, e.g. stray light, markings, dust, scratches, missing portions of the disc also hamper automatic calculation of disc centre coordinates and disc radius. 
This is unfortunate because errors in the determination of the disc centre and radius can contribute to miscalculating the plage areas.

The method we used is similar to that of \cite{ermolli_comparison_2009}, and includes the following processing steps.
\begin{itemize}
	\item We apply a Sobel filter \citep{duda_pattern_1973} on the image to roughly identify the edge of the solar disc. Sobel filter was found to be more efficient for this task than other similar filters. At this stage, the identification of the disc edge is still influenced strongly by the various image artefacts. 
	\item We create a mask by assigning the value of 1 to regions within appropriate thresholds in the image that resulted from the Sobel filter. 
	\item We dilate and erode the mask to make its edges smooth and then remove all small-size and isolated regions in the mask that are most likely due to image artefacts. 
	\item We apply the Sobel filter to the smoothed mask to identify the edge of the disc, creating yet another mask.
	\item We perform a bootstrap Monte Carlo simulation where we randomly choose 10$\%$ of the points from the mask and fit a circle. We repeat this process 1000 times and then adopt the median parameters for the radius and centre coordinates.
\end{itemize}

This method works well on all data, except when there are significant artefacts outside the disc, e.g. large scratches or stray light and exposure problems which can render the regions outside the disc to have similar density levels as the disc. 
In such cases, we defined the centre coordinates and the radius of the solar disc by fitting a circle to three manually selected points. The points were chosen to be roughly 120$^\circ$ apart. This allows an estimate of the disc centre and the radius with sufficient accuracy for further processing. 
If the disc shape was not circular then the points were chosen such that they include the largest part of the disc possible without including considerable parts of the regions outside the disc. This introduces bias, since a different selection of the points would result in different radius estimates and hence different plage areas. However, we should note that there were only a few such images for which we manually identified the radius. Furthermore, images with strongly distorted discs were ignored and should be in any similar study (see Fig. \ref{fig:01_01_shg_examplesexcluded2} d)).

It should be noted that in this process the radius is estimated as if the recorded solar disc was perfectly circular. This potentially introduces errors in our resulting plage areas for images with irregular or elliptical disc (see Sec. \ref{sec:sorting}). 
\cite{ermolli_comparison_2009} studied the eccentricity of the solar disc in the Ar, Ko, and MW datasets and found average values of 0.14, 0.12, and 0.12 respectively, with an increase of eccentricities over time for Ar and Ko data. We estimated that a disc eccentricity of 0.5 introduces an uncertainty of 0.01 in the fractional plage areas derived from observations taken at the activity maximum. 
For more information on this estimate see Sec. \ref{sec:eccentricities} of the appendix.

The analysed data show variable orientation of the solar disc, which is needed to be homogenised in order to produce a consistent Ca~II~K time series for certain applications. To ease comparisons, the images shown in Fig. \ref{fig:processedimagessamedayoriginal} and in later figures were rotated to place solar north at the top of the image and solar west at the right side of the image.
However, the results presented in the following were derived on images with variable orientation of the solar disc. The Ko, MW, and WS images have markings on the plates to denote the disc orientation. For Ko data, the markings are dots usually outside the solar disc, with two (one) dots denoting the north (south) pole; for MW data the markings have a rectangular (triangular) shape to denote the north (south) pole; for WS there is a line showing the pole axis and letter markings to denote the north direction. These markings were introduced either during the observation or just before the digitisation of the observation. For the images from these archives, we manually selected the markings on the images to get the angle needed to rotate the images to align the north at the top of the image. MM data seem to have been scanned in a very consistent way so that the orientation of the images can be simply derived by using the ephemeris. Ar data were taken with a coelostat, and a simple rotation of 90$^\circ$ was found to be sufficient. Mi and Sc data unfortunately lack any of those markings, while the orientation has changed considerably due to the digitisation. Considering that these archives often have long gaps, the only way to orient single images properly is to use one consistent archive as the reference and appropriately orient the other images. We used the data from MM as the reference and oriented the images from Sc and Mi to them. The same procedure was also applied to those images from the other archives that were found to have incorrect markings. For instance the pole markings for image Ko1 in Fig. \ref{fig:processedimagessamedayoriginal} are roughly 40$^\circ$ off, while a small correction was needed in Ar2 too.

Furthermore, the information as to which side corresponds to the east or west is contingent upon the way the plates were digitised and this is lost for the Ar, Ko, MW, and most Mi data. This is not an issue for data from WS which have letter markings to denote the west direction and data that were stored in a non-transparent medium, like those from MM or the latter data from Mi. Likewise with the north-south orientation, we used the data from MM to identify which side corresponds to east or west and appropriately orient the images from the other archives.
The east-west orientation was found to be random among the images shown in Fig. \ref{fig:processedimagessamedayoriginal}.

\section{Effects of disc ellipticity on plage areas}
\label{sec:eccentricities}
Here we describe our estimate of the uncertainty in our derived plage areas due to the distortions of the solar disc in the recorded images.
The digitised images have distorted discs due to uneven movement of the plate during the observation or the scanning device with a linear array (such as that used for the 8-bit Ko data) or even if the plate is not aligned properly with the digitising device. This renders the solar disc in the digital files more elliptical. An example of a distorted disc can be seen in \ref{fig:01_01_shg_examplesexcluded2}d). 
In order to estimate the error in the plage areas due to the disc ellipticity we made use of the Rome/PSPT data. We stretched the disc on the $x$-axis to mimic the distortions of the historical data. We considered 20 ellipticity cases in the range [0,0.54], while maintaining the same region to be considered as the identified solar disc. We note here that since the test was done on the Rome/PSPT data the accuracy of determining the disc is very high and by running the code on the stretched images resulted in the same estimate of the disc centre and radius as for the original images. However, this might not be the case for the historical data, where the presence of artefacts near the limb can affect the process and introduce further errors in our estimates of the centre and radius. 
Furthermore, we applied equal stretching for the entire $x$ dimension, which is not entirely consistent with the historical data which can have only a few strips stretched and most likely in an uneven manner. Nonetheless, this test helps to estimate the order of magnitude of the error.

\begin{figure}[t]
	\centering
	\includegraphics[width=1\linewidth]{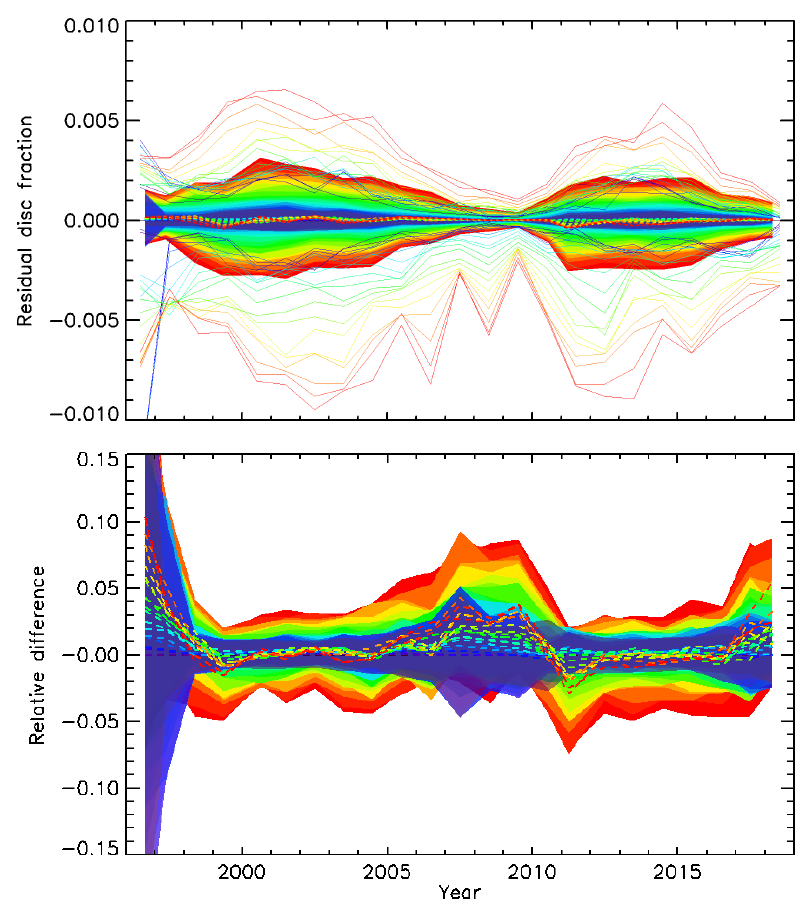}
	\caption{Absolute (top panel) and relative (bottom panel) difference of derived plage areas from images with imposed disc ellipticities to those from the original Rome/PSPT data. Shown are the annual median values (dashed lines), the asymmetric $1\sigma$ interval (shaded surfaces), and the annual maximum and minimum values (solid lines). The colours denote results for different disc eccentricities between 0.025 (dark blue) and 0.5 (red).}
	\label{fig:ellipticityerrors}
\end{figure}

Fig. \ref{fig:ellipticityerrors} shows the differences in the derived plage areas from the images with imposed ellipticities and the original Rome/PSPT ones. 
The scatter of the absolute differences increases during activity maxima and reaches roughly $^{+0.007}_{-0.01}$.
We notice that the annually averaged differences remain less than 0.001 even  for an ellipticity of 0.5. The relative differences reach values up to $\sim$5\% during activity maxima. During activity minima they reach up to 30\% (not shown in the figure). We note, however, that during low-activity periods disc fractions are often very close to zero, which leads to infinite relative errors. Therefore, we calculated the relative errors by considering only disc fractions greater than 0.001.

\section{Comparison of Meudon and Rome/PSPT data}
\begin{figure}
	\centering
	\includegraphics[width=1\linewidth]{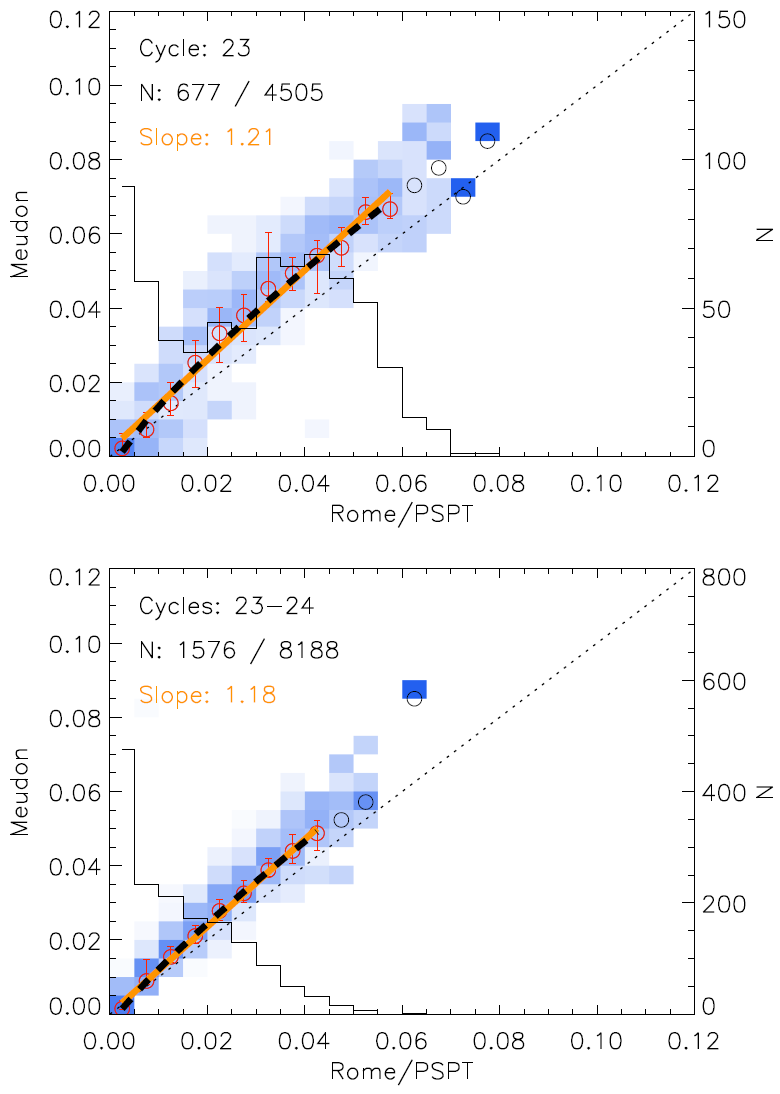}
	\caption{Same as Fig. \ref{fig:meothers} but now comparing Rome/PSPT and Me by taking Me as the reference, for the historical (top) and modern (bottom) Me observations separately.}
	\label{fig:mdpsptcycles14-25chihist}
\end{figure}

Figure \ref{fig:mdpsptcycles14-25chihist} shows the PDF matrices comparing Me and Rome/PSPT data as in Fig. \ref{fig:meothers}, but here for the historical and modern Me data separately.
The overlap between Rome/PSPT and modern Me data starts around the maximum of SC 23 and includes SC 24, while the overlap with the historical data covers only the ascending phase of SC 23 up to the maximum.
For modern Me data, the number of days common with Rome/PSPT is twofold the number for the historical Me data, yet the scatter is lower. It should be noted, however, that the historical data include periods with higher plage areas, which contributes, at least partly, to the higher scatter. Still, even for the historical data, the scatter is comparatively low.   
Overall, the relationships we find between Rome/PSPT on one side and modern or historical Me data on another one are very close. For the modern Me data alone, the slope of the linear fit is 1.18, while it is 1.21 if either only historical or all Me data are used. 

\end{document}